\documentclass[final,5p,times,authoryear]{elsarticle}

\usepackage{epsfig}

\usepackage{amssymb}

\usepackage{newtxtext,newtxmath}

\usepackage[T1]{fontenc}

\usepackage{subcaption}
\usepackage{multicol}
\usepackage{comment}
\usepackage[dvipsnames]{xcolor}

\newtheorem{definition}{Definition}[section]

\def\tx{ \textstyle} 

\usepackage{lastpage}

\usepackage{, bm}

\usepackage{hyperref}
\usepackage{graphicx}
\usepackage{amsmath}
\usepackage{float}
\usepackage{commath}
\usepackage{geometry}
\usepackage{tabularx}
\usepackage[toc,page]{appendix}
\usepackage[noadjust]{cite}
\usepackage{aas_macros}

\journal{Astronomy and Computing}

\begin{document}

\begin{frontmatter}

\title{\textbf{Graph Theoretical Analysis of local ultraluminous infrared galaxies and quasars}}

\author[Orestis Pavlou et al.]{Orestis Pavlou\corref{cor1}}

\cortext[cor1]{Corresponding author:}
\ead{o.pavlou@euc.ac.cy}
\author[Orestis Pavlou et al.]{Ioannis Michos}
\author[Orestis Pavlou et al.]{Vicky Papadopoulou Lesta}
\author[Orestis Pavlou et al.]{Michalis Papadopoulos}
\author[Orestis Pavlou et al.]{Evangelos S. Papaefthymiou}
\author[Orestis Pavlou et al.]{Andreas Efstathiou}

\affiliation[Orestis Pavlou et al.]{
            organization={School of Sciences, European University Cyprus},
            addressline={Diogenes Street, 6, Engomi}, 
            city={Nicosia},
            postcode={1516}, 
            country={Cyprus}}

\begin{abstract}
We present a methodological framework for studying galaxy evolution by utilizing Graph Theory and network analysis tools. We study the evolutionary processes of local ultraluminous infrared galaxies (ULIRGs) and quasars and the underlying physical processes, such as star formation and active galactic nucleus (AGN) activity, through the application of Graph Theoretical analysis tools. We extract, process and analyse mid-infrared spectra of local (z < 0.4) ULIRGs and quasars between 5-38$\mu m$ through internally developed Python routines, in order to generate similarity graphs, with the nodes representing ULIRGs being grouped together based on the similarity of their spectra. Additionally, we extract and compare physical features from the mid-IR spectra, such as the polycyclic aromatic hydrocarbons (PAHs) emission and silicate depth absorption features, as indicators of the presence of star-forming regions and obscuring dust, in order to understand the underlying physical mechanisms of each evolutionary stage of ULIRGs. Our analysis identifies five groups of local ULIRGs based on their mid-IR spectra, which is quite consistent with the well established \textit{fork} classification diagram by providing a  higher level classification. We demonstrate how graph clustering algorithms and network analysis tools can be utilized as unsupervised learning techniques for revealing direct or indirect relations between various galaxy properties and evolutionary stages, which provides an alternative methodology to previous works for classification in galaxy evolution. Additionally, our methodology compares the output of several graph clustering algorithms in order to demonstrate the best-performing Graph Theoretical tools for studying galaxy evolution. 

\end{abstract}


\end{frontmatter}

\section{Introduction}

 \subsection{Theoretical framework and Motivation}

One of the major fields of study at the forefront of astrophysical research is the study of galaxies in the observable Universe. Ultraluminous infrared galaxies (ULIRGs) are a specific type of galaxies with $1-1000 \mu m$ luminosities exceeding $10^{12} L_{\odot}$, discovered by the Infrared Astronomical Satellite (IRAS) in the 1980s \citep[e.g.][]{Soifer1987, Neugebauer1984}. They have since been the subject of several studies  \citep{Sanders1988,RowanEfstathiou93,Genzel1998,Farrah2003,imanishi07,vei09,Farrah2013}. ULIRGs are galaxies undergoing a merger event: when two or more gas-rich disk galaxies interact \citep{SandersMirabel96,Perez2021}, they gradually lose their orbital energy and angular momentum, collide with one another and ultimately merge into one larger galaxy. This process (called a \textit{merger}) results in an increase in their luminosity. \par

The most dominant evolutionary phases of ULIRGs in the merger scenario are the starburst-dominated phase and the AGN-dominated phase \citep{efstathiou95,Genzel1998,Rigopoulou1999,Sturm2000,Spoon2004,Spoon2007}. In this merger scenario, the observed increase in luminosity of these galaxies is caused by the gravitational interaction (and later merger) of two galaxies. During the \textit{pre-merger} phase, this interaction re-distributes stars, as well as the interstellar gas and dust present in each of the galaxies, changing the shape of the galaxies. As this happens, molecular clouds are formed and destroyed, causing new concentrations of interstellar gas and dust which lead to the formation of new active star formation regions within each galaxy. This sudden increase in star formation rate is called the starburst-dominated phase \citep{Efstathiou2000,Farrah2001,Farrah2003}. The following phase in the merger scenario is called the \textit{coalescence} phase, the intermediate phase where the two galaxies slowly coalesce into one larger galaxy, with a single galactic nucleus \citep{Genzel1998,Farrah2003,Farrah2009,Farrah2013}.  Recent models for galactic mergers have shown that at this stage one or both of the supermassive black holes (SMBHs) at the core of each galaxy display an increase in activity, as they begin to accrete matter. The phenomenon associated with an active SMBH at the center of a galaxy is called an \textit{active galactic nucleus} (AGN). The activation of an AGN is observed as a boost in the AGN fraction of the ULIRG's luminosity, meaning that the fraction of the overall luminosity which is attributed to the AGN increases. Finally, the ULIRGs pass into the \textit{post-merger} phase, where the merged galaxy becomes a luminous AGN and the starburst activity drops dramatically. This is caused by the disruption of the feedback process of the SMBH at the core of the galaxy due to increased AGN activity, with galactic winds and highly relativistic jets expelling matter from the core of the galaxy, `starving' the rest of the galaxy of excess dust and gas, while the re-distributed interstellar dust and gas has already been consumed by the starburst phase, during the formation of new stars. The physical processes behind this evolutionary scenario are what we aim to unravel by applying our proposed methodology, using graph theoretical analysis methods and algorithms. \par

Starburst galaxies display different characteristics than galaxies that go through an AGN-dominated phase. The starburst phase means that a galaxy has an excess of gas and dust, which collapse under gravity to form new stars at a very high rate, whereas the AGN-dominated phase includes the accretion of matter onto the galaxy's SMBH, which generates large amounts of energy and results in the ejection of gas and dust from the galaxy, via highly energetic jets. Usually this source of energy is obscured by the AGN torus - a dusty toroidal structure surrounding the central SMBH at the center of each galaxy, as predicted by the unified model of AGNs \citep{Antonucci1993}. This process results in the formation of quasi-stellar objects (QSOs, commonly known as quasars). Quasars are thought to be complementary objects to ULIRGs, but some studies challenge this single evolutionary scenario and suggest the existence of multiple different evolutionary paths to be undertaken by ULIRGs \citep[see ][and references therein]{Farrah2003}. Interestingly, we note that recent studies \citep{Kirkpatrick2020} detected a population of cold quasars which are not purely AGN objects, as they seem also to be accompanied by strong star formation activity. Strong star formation activity was also inferred in the quasar samples studied by \citet{Harris2016} and \citet{Pitchford2016}. The effects of the underlying mechanisms and processes of the merger scenario result in certain differences in the observed characteristics of ULIRGs and quasars. Studying these differences is the key to understanding the evolutionary stages these galaxies go through. \par

Star formation dominates the far-infrared emission in the spectra of ULIRGs, whereas AGN can make a significant contribution and even dominate the emission of ULIRGs at near- and mid-infrared wavelengths \citep{Farrah2003,Marshall2007, Nardini2008, Vega08,Efstathiou2022}. Extracting the contributions of star formation and AGN activity remains a major challenge. This is mainly due to the presence of dust that hides the energy sources of these galaxies, which results to an absorption of their  energetic radiation. Eventually, the absorbed energy is re-emitted as infrared thermal radiation. Understanding  local ULIRGs constitutes an important step towards the interpretation of  submillimeter galaxies (SMGs) \citep[see][]{Hughes98,Barger1998,Casey2014} and other populations of galaxies, where extreme starburst and AGN activity occurred in the history of the Universe \citep[e.g.][]{Harris2016, Pitchford2016,Rowan-Robinson2018}. \par

Galaxy spectra, from the ultraviolet to the millimetre (0.1-1000$\mu m$), encode information about their nature and their evolutionary state. Models for their emission \citep{efstathiou95,Silva1998,Efstathiou2000,Siebenmorgen2007,efstathiou09} show that medium resolution spectroscopic data in the 5-35$\mu m$ wavelength range are required for determining the properties of galaxies such as uncovering obscured AGN, powered by accretion onto SMBHs and determining the age of the starburst episode. In this work we study the evolution of galaxies by examining mid-infrared spectra of local ultraluminous infrared galaxies and quasars (at $z<0.4$) with the use of Graph Theory and network analysis.  \par

The observed characteristics in the mid-infrared spectra of galaxies which have been proven to be indicators for the presence of high star formation rate are the emission features of polycyclic aromatic hydrocarbons (PAHs) in specific mid-IR wavelengths ($6.2, 7.7, 8.6, 11.2, 12.7 \mu m$), \citep{Roche91,Genzel1998,Peeters2002}. The mid-infrared spectrum also contains information of emission and absorption features around 9.7 and 18$\mu m$ caused by the presence of silicate dust. The silicate emission features arise due to the emission of hot dust at a temperature of $\sim $300-1000K which is viewed almost completely unobscured when the AGN tori are viewed face-on \citep{Rowan-Robinson1993,GranatoDanese1994,efstathiou95}. The absorption features at 9.7$\mu m$ are more difficult to interpret, as they may be either caused by the viewing angle of the torus being edge-on (buried AGN) \citep{Pier1992,GranatoDanese1994,efstathiou95,imanishi07} or by obscured star formation \citep{RowanEfstathiou93}. \par

\citet{Murata2016} studied the relationship between PAH emission and star formation in nearby galaxies at $z < 0.2$, by analyzing data on 55 star-forming galaxies from different instruments (AKARI, WISE, IRAS, Hubble Space Telescope and SDSS) and found that PAHs are partially extinguished during the late stages of galactic mergers. The most likely culprits for this are strong radiation fields and large-scale shocks taking place during mergers. In a separate work by \citet{Shipley2016}, the authors study the PAH emission features of galaxies at $1$<$z$<$3$ to calibrate their star formation rates (SFRs). Using Spitzer (IRS) observations they demonstrate how PAH emissions can accurately describe SFRs in galaxies and thus help distinguish between star-forming and AGN-dominated galaxies. The classification of the evolutionary phase of galaxies via the study of silicate absorption strength and PAH emission in the mid-infrared range was also performed by \citet{Spoon2007}. The authors of this work show how the specific emission features of PAHs at 6.2$\mu m$ and silicate strength at 9.7$\mu m$ can be a useful indicator to distinguish between starburst and AGN dominated galaxies. They present their results graphically in a `Fork' classification diagram. \par

Recent studies of luminous infrared galaxies have revealed differences in the properties of these galaxies at low vs high redshift. The most luminous (ultraluminous and hyperluminous) infrared galaxies at low redshift are consistent with a galactic merger scenario. However, studies with Herschel \citep{Ridighiero2010} discovered the lesser role of starbursts triggered by mergers in the high redshift Universe. The idea that `Main Sequence' galaxies, or galaxies with significant contribution from `cirrus', can reach infrared luminosities exceeding $10^{12} L_{\odot}$ was also suggested pre-Herschel by \citet{Efstathiou2003} and \citet{efstathiou09}. Intense obscured starburst and AGN activity triggered by mergers is still considered to be the dominant power source in the most luminous galaxies observed in Herschel surveys \citep{Rowan-Robinson2018,Gao2021}. Current ideas also suggest that the mergers that occurred in the early Universe may not have been as efficient at triggering starbursts as similar events in the local Universe. Thus, the local merger scenario may not cover the entire sample of observed luminous galaxies in the early Universe (e.g. \citep{Schreiber(2015)}). \par

The more traditional study of spectral energy distributions (SEDs) of galaxies involves fitting their spectra with models which incorporate stellar population synthesis models \citep{BruzualCharlot1993,BruzualCharlot2003} and treat the effects of dust either in a simplified geometry (e.g. in codes like MAGPHYS and CIGALE) or in more detailed treatments like radiative transfer models like GRASIL \citep{RowanEfstathiou93,Silva1998,Vega08,efstathiou13,Efstathiou2014,Efstathiou2021,Efstathiou2022}. In addition, an AGN model is usually included in the fitting and in some cases the AGN can dominate the luminosity. The approach adopted by \citet{Farrah2009} using Graph theory and Bayesian inferencing to study the evolution of 102 local ULIRGs is an alternative to the SED fitting approach and it has the advantage that it can lead to the identification of different phases in the evolution of luminous infrared galaxies and the development of an evolutionary paradigm. \citet{Farrah2009} successfully developed an evolutionary paradigm for low redshift infrared galaxies using this approach. \par

In this work we analyze data obtained from NASA's \textit{Spitzer Space Telescope} and in particular the \textit{Infrared Spectrograph (IRS)} instrument onboard that spacecraft. More specifically, we extract, process and compare mid-infrared wavelength ($5-35 \mu m$) spectra of local ($z<0.4$) ultraluminous infrared galaxies (ULIRGs) and quasars, in order to generate a relational (similarity) graph of these galaxies and use Graph Theoretical tools to perform network analysis. We use publicly available data provided via the \textit{CASSIS} website (\href{https://cassis.sirtf.com/atlas/welcome.shtml}{Combined Atlas of Sources with Spitzer IRS Spectra}) of infrared galaxy sources. There are several important samples consisting of hundreds of galaxies \citep{Fu2010,Sajina2012,Kirkpatrick2015,Kirkpatrick2020} with publicly available data through the CASSIS database (see \citet{Lebouteiller2011,Lebouteiller2015}). As in the case of the sample studied by \citet{Farrah2009}, the samples include mid-infrared data from the Spitzer Space Telescope which are essential for this kind of analysis. This approach is especially useful for studying the evolution of infrared galaxies given the availability of data for large samples of galaxies that are already available (e.g. from projects like HELP \citet{Shirley2019,Shirley2021}) and which will soon be complemented with spectroscopic data from the James Webb Space Telescope (JWST) which was launched by NASA in December 2021. \par 

\subsection{Graph Theory in Astrophysics}

Graph Theory is the discipline of Mathematics and Computer Science involved with the study of the relations between objects, based on their relational properties, which are extracted from the data into matrices. These matrices are then used to generate and map the objects in a network structure called a \textit{graph}. The objects are presented as \textit{nodes} (vertices) in the graph, and are connected via \textit{edges}, which correspond to their relational properties. One of the major approaches to extract meaningful information from a graph, is the application of \textit{clustering algorithms}, which are unsupervised learning techniques that separate the graph into distinct groups (clusters) of nodes which are more highly connected between them compared to the rest of the graph. \par

In our specific case, we apply a \textit{similarity function} to the mid-infrared spectra of ULIRGs in order to create a \textit{similarity matrix}. This similarity matrix is then used to generate our graph, which is called a \textit{similarity graph}. In this similarity graph, the nodes represent ULIRGs, which are connected based on the similarity of their mid-infrared spectra. We then apply, test and compare several clustering algorithms, resulting in the detection of communities of ULIRGs with similar spectra. \par

One of the main advantages of this approach is the fact that it not model-dependent, as the entire process of generating the similarity graph is based purely on the pairwise similarities between nodes obtained by comparing the corresponding observational data. Therefore, the implementation of this methodological approach to astrophysical data, and in particular to the mid-infrared spectra of ULIRGs, is completely independent of any model parameters (compared to model fitting approaches which are heavily dependent on model parameters). Thus, the application of graph theoretical methods for the study of the merger scenario of ULIRGs can produce model-free results, by using the clustering of ULIRGs into similar evolutionary stages based on the similarities between their mid-infrared spectra, in order to generate an evolutionary paradigm for ULIRGs. \par

Despite the  long history of graph theory and its successful implementation in various sciences, there are very few works that exploit Graph Theory in order to study open problems in Astrophysics and specifically galaxy evolution. The only major work that utilized Graph Theory for the study of ULIRG evolution was performed by \citet{Farrah2009}. The authors study mid-infrared spectra of a set of Spitzer (IRS)  ULIRGs by combining the methods of Graph Theory and Bayesian inferencing in an attempt to identify and distinguish between different phases of temporal evolution. They select a local (z $<$ 0.4) ULIRG population and utilize Bayesian inferencing on their mid-infrared spectra to produce a {\em similarity} graph, where galaxies of similar SEDs are connected to each other in the graph. Then, using a spring-layout \citep{Kobourov2012} drawing of the graph (where similar nodes are placed closed in the graph embedding), they identify 3 groups of galaxies which are suggested to correspond to different evolutionary stages of the galaxies.
 
Although the exploration of graph theory in astrophysics is limited, more research works have been performed in the field of cosmology. For example, \citet{Coutinho(2016)} utilized dynamical network analysis of cosmological models of large-scale structures, containing a number of simulated galaxy distributions, in order to study the gravitational interactions and evolution of galaxy clusters and superclusters. Additionally, \citet{Hong(2015)} as well as \citet{Sabiu(2019)} showcased how graph theoretical methods and tools can be successfully applied to simulated as well as observational data. The aforementioned works suggest that the application of Graph Theory can also be a very useful tool in the cosmological study of galaxy formation and distribution. 

Finally, the recent papers of \citet{papaefthymiou2022} and \citet{frontiers_paper2022} have demonstrated uniform frameworks for performing analysis and classification of ultraluminous infrared galaxies based on Graph Theory and the utilization of graph clustering tools. The applicability of the framework is demonstrated in diverse application domains, which include astrophysical data analysis to study galaxy evolution.

\subsection{Methodology overview}

In this paper we propose a new classification diagram for ULIRGs and quasars, derived from graph theoretical analysis; in particular graph clustering. This approach enables us to group and classify the spectra of the ULIRG sample of \citet{Farrah2009} along with the spectra of 37 Palomar Green quasars \citep{Symeonidis2016}.

Our  methodology consists of three steps: ({\em i}) We model the data with a certain type of a weighted graph (a {\em similarity graph}), in which nodes represent galaxies and edges are weighted based on the similarity  of their end-points. More precisely, weights measure the similarity between the SEDs of galaxies.  ({\em ii}) We apply graph clustering algorithms to identify groups of similar galaxies (clusters) and use well-known graph theoretical measures to validate the clustering. ({\em iii}) Finally, we interpret each cluster from an astrophysical point of view, in terms of the physical properties of ULIRGs, such as the PAH emission and silicate absorption features.

The proposed methodology results in a classification scheme that: ({\em a}) improves the grouping in \citet{Farrah2009}, since it employs the application of graph clustering, and ({\em b}) generalizes the popular diagram of \citet{Spoon2007}, by viewing the resulting clustering in the PAH emission - Silicate absorption plane.

\noindent {\bf Roadmap:} The paper is organized as follows: In section \ref{Data Description} we describe the observational data we used in our analysis. We use the spectroscopic data obtained with the IRS instrument on-board Spitzer \citep{houck04} which covers the wavelength range $5-35\mu m$ \citep{Lebouteiller2011}. In section \ref{Mathematical Framework}, in order to be self-contained, we briefly describe the Kernel PCA and graph theoretic methodology we employ. In section \ref{Methodology_Section} we describe our methodology for the construction of the graphs, as well as the graph clustering algorithms' implementation in more detail. In section \ref{ResultsSection} we present and discuss our results. Finally, in section \ref{Conclusions} we present our conclusions. \par

Throughout this work, we assume a spatially flat Universe, with \mbox{$H_0 = 70$\,km\,s$^{-1}$\,Mpc$^{-1}$}, \mbox{$\Omega = 1$}, and \mbox{$\Omega_{\Lambda} = 0.7$}.

\section{Data Description}\label{Data Description}

\subsection{Observational Data}

The sample of \citet{Farrah2009} consists of 102 local ULIRGs with redshifts between $0<z<0.4$. The data were acquired by NASA's Spitzer Space Telescope's \citep{Werner2004} Infrared Spectrograph (IRS) instrument \citep{houck04}, which covers the wavelength range between $5-35\mu m$. The spectra were extracted from the \textit{Combined Atlas of Sources with Spitzer IRS Spectra}\footnote{The data is publicly available at the CASSIS website: \url{https://cassis.sirtf.com/}} (CASSIS) website, with ID:105 - \textit{Spectroscopic Study of Distant ULIRGs II} by James R. Houck \& Lee Armus. This dataset is comprised of 118 objects observed in Low-Resolution and 53 observed in High-Resolution. Similarly to \citet{Farrah2009}, due to the lack of any synchronous sky-background data for the High-Resolution set, we chose to study the Low-Resolution dataset of CASSIS version LR7, released in June 2015 by \citet{Lebouteiller2015}, which is a more mature pipeline version of the data than the one used in \citet{Farrah2009}, originally presented in \citet{Desai2007}.

Palomar Green (PG) quasar spectra were added to our sample, taken from \citet{Symeonidis2016}, in order to create a more complete total sample of local ULIRGs and quasars. Due to the limited number of quasars present in the previous sample (in \citet{Farrah2009}), we decided to make this addition to generate a more significant sample of quasars within our total sample. This was necessary in order to eliminate any potential sample bias caused by the much higher number of ULIRGs in other evolutionary phases (i.e. starburst-dominated) influencing the clustering analysis of our similarity graph. We were able to extract IRS spectra from the CASSIS website for $58$ PG quasars. Due to the fact that some of these quasars were only observed between $14-37\mu m$, a spectral region that excludes the range of interest studied in this work, namely certain PAH emission and silicate features, we had to limit our sample size to 42 quasars. By cross-referencing this sample with the ULIRG sample of \citet{Farrah2009}, we were able to add 37 of the aforementioned PG quasars, with redshifts between $0<z<0.2$ to the ULIRG sample, bringing our total sample size to 139 ULIRGs and PG quasars.

\subsection{Data pre-processing}

The data pre-processing was accomplished completely via our own purpose-made Python code for extracting and pre-processing the aforementioned data \footnote{Our purpose-developed Python code for feature extraction, processing routines and application of our graph theoretical analysis and algorithms are available in a public \href{https://github.com/orestispavlou/Graph-Theoretical-Analysis-of-Ultraluminous-Infrared-Galaxies-and-Quasars-Pavlou-et-al.-2023-.git}{Github repository}.}. Firstly, we extracted, cross-referenced and corrected the redshift values of each galaxy using the \textit{NASA/IPAC Extragalactic Database (NED)}\footnote{NASA/IPAC Extragalactic Database (NED) website: \url{https://ned.ipac.caltech.edu/}}. After filtering the spectra to remove possible double flux values, caused by overlapping orders between the Short-Low (SL) and Long-Low (LL) modules of the instrument for specific wavelengths \citep{houck04}, we calculated the rest wavelengths: $\lambda_{rest} = \frac{\lambda_{obs}}{1 + z}$ and interpolated flux densities per rest wavelength bin, between $5.5 - 26.0 \mu m$.

The next steps of our Python pre-processing routines involve continuum fitting to the data to remove the continuum from the observed signal, in order to estimate more accurately certain important physical features present in infrared (rest) wavelengths between $6.2 - 12.7 \mu m$. In this wavelength range, we calculate the emission signals of polycyclic aromatic hydrocarbons (PAH) at $6.2, 7.7, 8.6, 11.2$ and $12.7 \mu m$ and the absorption/emission feature of silicate dust particles at $9.7\mu m$.

\subsection{Feature extraction}

As mentioned previously, these PAH features are indicators of the rate of star formation activity in a galaxy. As their name suggests, they are chemical compounds made up of hydrogen and carbon atoms and they have been shown to be abundant in cold molecular clouds \citep{Lebouteiller2011}. These molecular clouds have been extensively studied in several galaxies, and are mostly present in regions of active star formation (also referred to as `stellar nurseries'), where young stars form at an increased rate compared to the rest of the galaxy.

For the calculation of the PAH equivalent width, we wrote our own Python routine which interpolates the flux values at each PAH wavelength (i.e. $6.2\mu m$) and then makes use of scipy's \textit{InterpolatedUnivariateSpline}\footnote{Scipy - Interpolated Univariate Spline tool: \url{https://docs.scipy.org/doc/scipy/reference/generated/scipy.interpolate.InterpolatedUnivariateSpline.html}}, in order to integrate between values $\pm0.3\mu m$ around each wavelength (i.e. between $5.9\mu m$ - $6.5\mu m$ for the $6.2\mu m$ PAH feature). We then use this integral result ($F_{\lambda}$) to calculate the equivalent width ($W_{\lambda}$) as:

\begin{center}
    {$W_{\lambda} = \int(1 - F_{\lambda}/F_c) d \lambda$}
\end{center}

where $F_c$ denotes the flux of the continuum, estimated by using the flux values $\pm0.3\mu m$ around each PAH wavelength. \par

For example, the calculation used for the $6.2\mu m$ PAH feature will be:  
\begin{center}
    {$W_{6.2} = \int_{5.9}^{6.5} (1 - F_{6.2}/F_{5.9-6.5}) (0.6)$}
\end{center}

Similarly, as mentioned previously, the \textit{silicate feature} at $9.7\mu m$ has been linked to the presence of silicate dust in the line-of-sight between the observer and the AGN in the core of ULIRGs, mainly in an area around the AGN called the \textit{torus}. Silicate dust can exhibit either an emission or an absorption feature, unlike PAHs. For the calculation of the silicate feature strength, we use the \textit{specutils} tools\footnote{Specutils tool (Astopy) documentation: \url{https://specutils.readthedocs.io/en/stable/}} of the \textit{Astropy} Python package, to create a spectrum from the flux density, perform continuum fitting using the \textit{Chebyshev1D} model and the \textit{LinearLSQFitter} module and then divide the flux over the fitted continuum. This is a necessary step before calculating the silicate feature strength, as the feature is quite broad and is prone to the distortions of the spectrum, due to continuum differences between ULIRGs. This is a similar technique to the one used by \citet{Spoon2007}. \par

The silicate absorption/emission feature strength calculation formula is:

\begin{center}
    {$Si = \log ( \frac{F_{obs}}{F_{cont}}$} )
\end{center}

where $F_{obs} =$ observed flux at $9.7\mu m$ and $F_{cont} =$ continuum flux at $9.7\mu m$. \par

The calculated PAH emission equivalent widths for $6.2\mu m$ and $11.2\mu m$ and the silicate absorption strengths at $9.7\mu m$ based on the aforementioned methodology implemented in our work are presented in Table \ref{PAH_Si_table} in \ref{appendixIV}. It is worth noting that these results show some level of consistency compared to the corresponding values presented in other works (e.g. \citet{Desai2007}, \citet{Spoon2007}, \citet{Farrah2009}). There are some deviations in the estimation of these values compared to those works for certain specific cases of ULIRGs, but the overall trend of the data is in agreement with these works. This is apparent when comparing our versions of the \textit{fork} diagrams (shown in Figure \ref{Spoon_Shi_Malik} of this work) to the fork diagram of \citet{Spoon2007}, where most of the ULIRGs are consistently placed within the same regions of the nine diagram regions (1A-3C), as discussed in Section \ref{fork_diagrams} of this work. \par

\section{Mathematical Framework}\label{Mathematical Framework}
In this section we present the graph theoretical framework that is utilized in this work, which includes the similarity functions and graphs and graph clustering algorithms and quality measurements.

\subsection{Graph Theory Background}\label{GraphTheorySection}
Graph Theory, one of the oldest branches of Mathematics, is undeniably one of the most powerful tools of Applied Mathematics with remarkable interdisciplinary applicability in diverse areas, spanning from Social and Political Sciences, to Biology, Chemistry, Neuroscience \citep{Bullmore2009} and Astrophysics and Cosmology \citep{Ueda2003,Farrah2009,Hong2016} and stands out as a foundational concept of Network Science \citep{Barabasi2016}.

Graphs are mathematical structures that represent any type of entities which can be related to each other via pairwise relations. Each entity is represented by a vertex (node) of the graph. Pairwise relations between entities are represented via edges (links) connecting pairs of corresponding nodes. Additionally, a numerical value (weight) can be associated with each edge, which acts as an indicator of the strength of the relation between the corresponding nodes. For similarity graphs in particular, weights quantify the degree of {\em similarity} between the endpoints of their edges \citep{Blondel2004}.

More formally, an un-directed, weighted and loop-less {\em graph} $G$ is an ordered triple $G = G(V, E, w)$, where $V = [n] = \{ 1, 2, \ldots, n \}$ is the vertex (or nodes) set; $E \subseteq \{ (i, j ) \, : i,j \in V, \ i\neq j \}$ is the edge set; and $w$ is a {\em weight function} $w : E \rightarrow  \mathbb{R^{+}}$, specifying a positive {\em weight} $w(i,j)$ for each edge $(i, j ) \in E$. For any fixed vertex $i \in V$,  the {\em (weighted) degree} of $i$ is defined as $\displaystyle {{\sf deg}_i \equiv \sum_{(i,j) \in E} w(i, j)}$. The {\em degree vector} of $G$ is the column vector $\mathbf{deg} = ({\sf deg}_1, \ldots, {\sf deg}_n)^{ T }$ and the corresponding {\em degree matrix} of $G$, is the diagonal matrix $\mathbf{D} = {\rm Diag} ({\sf deg}_1, \ldots, {\sf deg}_n)$.

A concise way to represent $G$ is through its \textit{(weighted) adjacency matrix}: an $n \times n$ matrix $\mathbf{A} = (a_{ij})$, where $a_{ij} \equiv w(i,j)$, for each $(i,j) \in E$ and $a_{ij} \equiv 0$ otherwise. 
Note that for un-directed graphs, the adjacency matrix is {\em symmetric}, and consequently all its eigenvalues are real numbers. 

Another useful matrix is the {\em (discrete) Laplacian} matrix, which - when viewed as a linear operator - is a discrete analogue of the Laplace operator $\bigtriangleup f = {\bigtriangledown}^2 f$ of a real-valued function $f$. It relates to many useful geometric properties of a graph and is defined as $\mathbf{L} \equiv \mathbf{D} - \mathbf{A}$. To be able to compare such properties among graphs of varying order, one has to consider certain normalized versions of ${\mathbf{L}}$, since $\mathbf{L}$ is un-normalized. The most common ones in the literature are: the {\em normalized symmetric Laplacian} ${\mathbf{L}}_{sym} \equiv {\mathbf{D}}^{-1/2}{\mathbf{L}}{\mathbf{D}}^{-1/2}$ and the {\em normalized random walk Laplacian}  ${\mathbf{L}}_{rw} \equiv {\mathbf{D}}^{-1}{\mathbf{L}}$. 
All versions of the Laplacian are symmetric and {\em positive semi-definite} matrices, hence their eigenvalues are real non-negative numbers. Furthermore, the eigenvalues of the normalized Laplacians are bounded in the interval $[0, 2]$, regardless of the order of the graph into consideration.

\noindent {\bf Graph Properties:} For any $S \subseteq V$, we denote by $\overline{S}$ its complement $V \backslash S$ in $V$.
%Consider a weighted graph $G = G(V, E, w)$. 
Two vertices $i, j\in V$ are considered \textit{neighbours} if there is an edge connecting them, i.e. $( i,j ) \in E$. 
The {\em Volume} of  $S \subseteq V$ is the sum of the degrees of its vertices, i.e., $Vol(S) \equiv \sum_{i \in S} {\sf deg}_i$.  We also define the volume of the whole graph to be $Vol(G) \equiv  Vol(V)$.  If $e$ is the sum of all weights of the edges of $G$, then clearly $Vol(G) = 2e$.
A \textit{path} in $G$ is a sequence of nodes $\{ v_1, \ldots, v_i, v_{i+1}, \ldots, v_k\}$ such that $( v_i, v_{i+1} ) \in E$.  A graph is \textit{connected} if there is at least one path between any pair of vertices. A graph $G$, in which   $\forall  i, j \in V$, it holds that   $(i,v)\in E$, is called a {\em complete} graph.
A graph $G'=(V',E')$ is a sub-graph of a graph $G=(V,E)$ if $V'\subseteq V$ and $E'\subseteq E$. If $G'$ contains all edges that join vertices $V'$ in $G$, we say that $G'$ is an induced sub-graph of $G$.

\subsection{Similarity Functions and Similarity Graphs}\label{Similarity Functions_Sect}

A {\em similarity function} ({\em similarity measure}) ${\sf} s$ is a function that uses information from entities $i,j$ and yields a real valued number ${\sf s}(i, j)$, with large values corresponding to high degree of similarity. Quantifying the similarity between all pairs of $n$ entities in a data set results in an $n \times n$ matrix, the {\em similarity matrix} $\ \mathbf{S} = (s_{ij})$, with $s_{ij} = {\sf s}(i, j)$. A specific similarity matrix $\mathbf{S}$ can be used as the adjacency matrix of a graph. The graph is weighted, with the weight function being the similarity function $s$; un-directed, since ${\sf s}(i, j) = {\sf s}(j, i)$;  and loop-less, after assigning 0 to all diagonal elements of $\mathbf{S}$. The corresponding graph is called {\em similarity  graph} and its adjacency matrix is called {\em similarity  matrix}. For the purposes of this work, we only consider similarity graphs.  

Most similarity functions use a distance metric (e.g. Euclidean distance) to compare different data-points. It's clear that such metrics actually quantify the dissimilarity rather than the similarity of the data. Turning dissimilarity into similarity can be done in different ways (e.g. negating, taking the inverse or using a negative exponent of a distance measure). One of the most popular choices, which is used throughout this work, is the {\em Gaussian kernel function}, commonly used in spectral graph clustering. A more detailed presentation of the Gaussian kernel function is presented in Section \ref{Similarity_Section}.

\subsection{The Graph Clustering Problem and its Quality measures}\label{GraphClusteringSection}

{\em Clustering} constitutes a fundamental problem in sciences: given a set (or system) of (any) discrete simple or structural data, we need to create a much smaller set (than the size of the actual data set) of groups of data, where each data point belongs to one of these groups and the data assigned to the same group have some significant relation (e.g. commonalities, dependencies) to each other.
Given a graph, the {\em  graph clustering} problem concerns the task of partitioning the nodes of the graph into groups, called {\em clusters} or
{\em communities}, such that the nodes within each group are highly connected
to each other and while the inter-crossing connections between nodes of different groups are as few as possible \citep{Fortunato2010}. Communities
indicate groups of nodes of the graph which are highly related to each other according to the modeling of the system through a graph.
In this work, we consider similarity graphs. A clustering in such a graph reveals partitioning of the nodes of the graph into nodes of similar properties, under the similarity property captured by the edges of the graph.
Although, there is no universal definition for the graph clustering problem, a   general, informal definition of graph clustering is the following:

\begin{definition}\label{clusteringDef} \normalfont
A \textit{graph clustering} of a weighted graph defined as $G(V, E, w)$ is a partition ${\cal C} = \{ C_1, \ldots, C_r \}$ of the vertex set $V$  into $r$ {\em clusters} or {\em communities}, for some $r \in \mathbb{N} $, i.e., $ C_i \subset V$, $C_i  \cap C_j = \emptyset$, for all $ C_i, C_j \in {\cal C}$ and $\bigcup_{i=1}^{r} C_i   = V$, such that the vertices within each cluster should be highly connected to each other, while the inter-crossing connections between vertices of different clusters should be as few as possible. 
\end{definition}

\subsubsection{Quality measures of Graph clustering} 
 
For a weighted graph $G$ and a subset $S \subseteq V$, let $e_{in}(S)$ denote the sum of the weights of all edges joining internal vertices of $S$, and $e_{out}(S)$ denote the sum of the weights of all edges with one vertex in $S$ and the other vertex in $\overline{S}$. Set also $e(S) = e_{in}(S) + e_{out}(S)$ and $e = e(V)$, when $S = V$. Note that clearly $vol(S) = 2 e(S)$. 

Using these notions we can express graph clustering as an optimization problem of computing a graph partition ${\cal C}$ in which for each cluster $C\in {\cal C}$, the ratio  $\frac{\tx e_{in} (C) } {  e_{out}(C)}$ is as large as possible. 
The following two notions (modularity and conductance) are quantitative measures for the performance of the graph clustering.

{\bf Modularity} 
The most known measure of a good graph clustering ${\cal C}$ is the notion of {\it{modularity}} which was first introduced by \citet{Girvan2002}. The modularity $m_{\cal C}$ measures the fraction of the edges that fall within clusters minus the expected number in an equivalent network, where edges were distributed at random, while the expected degree of the vertices is the same as the original graph. More formally:
$ m_{\cal C} \equiv \frac{1}{2e} \sum_{i, j\in V } \Bigl ( w_{ij} - {\frac{{\sf  deg}_{i}{\sf  deg}_{j}}{2e}} \Bigr ) {\delta}(C_{i}, C_{j}),  $
where ${\delta}$ is the {\em Kronecker delta function}, which in our case means that ${\delta}(C_{i}, C_{j}) = 1$ when the vertex $j$ lies in the cluster $C_i$ and ${\delta}(C_{i}, C_{j}) = 0$, otherwise.
One can confirm that $m_C \in [-1/2,1)$. Given a larger than expected portion of connections within a cluster, one can infer the presence of an underlying cluster structure. 
Finding a graph partition that maximizes modularity is considered NP-complete \citep{Brandes2008}.

{\bf Graph Conductance}   

Let $G$ be a connected weighted graph of $n$ vertices. For a subset $S \subseteq V$, the {\em conductance} of $S$ is defined as ${\phi}(S) \equiv \frac{e_{out}(S)}{e(S)}$. Since clearly ${\phi}(S) \in [0, 1]$, the conductance may be viewed as a probability measure, e.g., if ${\phi}(S) = 0.1$ it means that $90\%$ the neighbors of a random vertex in $S$ are expected to be in $S$. 

The conductance of a clustering of $r\geq 2$  clusters (or  $r$-way clustering)  $ {\cal C } =  \{ C_1, \ldots, C_i, \ldots, C_r \}$  is then defined to be the maximum (worst) of all the corresponding conductances of the clusters $C_i\in \cal C $.
The optimal $r$-{\em way conductance} for the graph $G$ is then defined as the minimum of the conductances of all possible $r$-way clusterings ${\cal C}$ of $G$.

\subsubsection{Graph Clustering Algorithms}

Graph clustering algorithms are essentially unsupervised learning techniques for the detection of communities (clusters) within a graph. There is an extensive literature on graph clustering algorithms with a multitude of approaches for the computation of clusters in graphs: hierarchically   \citep{Lian2007}, agglomeratively \citep{Newman2004b} or by divisive methods (e.g. \citet{Girvan2002,Newman2004PhysRevE}), driven by various measurements of clustering quality (e.g. modularity based \citet{Newman2004PhysRevE,Louvain}), using various graph properties, such as edges or cycles (e.g. \citet{Radicchi2004}), utilizing spectral graph theory and eigenvectors (e.g., \citep{Newman2006,Shi_Malik2000}), or ones that combine the aforementioned  approaches.

In this work we explore two of these important research approaches: ({\em  1})  Graph clustering algorithms applied directly on the similarity graph, which utilize modularity or conductance as a quality measure, and ({\em  2}) spectral graph clustering algorithms that utilize eigenvectors for performing the clustering and conductance to measure the quality of the clusters obtained. 
 
Within the first class of algorithms we explore (i.e. modularity-based algorithms), we investigate two of the most popular {\em hierarchical  agglomerative} algorithms:
The Leiden algorithm of \citet{Leiden}, which is an improved version of the Louvain algorithm of \citet{Louvain}, and the Walktrap algorithm of \citet{Pascal2006}. {\em Hierarchical agglomerative} clustering algorithms detect network communities by building a hierarchy of clusters, starting from  each node being a cluster on its own and progressively merging pairs of clusters until only one final cluster containing all nodes in the network is achieved, by trying to maximize the quality of the obtained clusters. Both algorithms explored in this work utilize modularity to find optimized clusters, as we explain below. Additionally, we also implement the Walktrap algorithm which uses conductance as an optimization measure. 
 
The Louvain algorithm  applies a greedy step approach in order to maximize the modularity of the clustering being constructed.
First, small communities are found by optimizing modularity through one-node alternations between neighbouring communities, until no further improvement of modularity is possible. Then each small community is grouped into one node and the first step is repeated. Here, we also apply the Leiden algorithm of \citet{Leiden}, which was suggested as an improvement of the Louvain algorithm, that yields communities that are guaranteed to be connected. The Leiden and Louvain algorithms are two of the most popular algorithms for community detection due to the fact that they perform very well in practice, both in terms of time efficiency and quality of the communities detected. 
  
The Walktrap algorithm of \citet{Pascal2006} is a hierarchical clustering algorithm based on random walks. The algorithm uses the notion of random walks to measure the similarity between two vertices, exploiting the observation that "random walks on a graph tend to get “trapped” into densely connected parts corresponding to communities" \citep{Pascal2006}.
The corresponding {\em distance measure} matrix obtained, is then iteratively used for aggregating nodes (and groups) of nodes into clusters trying to maximize a clustering quality measure (e.g. modularity or conductance). A clustering of optimal quality (in the level of aggregation) is returned.

The second kind of graph clustering algorithms we explore for this work are spectral graph clustering algorithms.  
These algorithms use the spectrum (eigenvalues and eigenvectors) on a similarity matrix (associated with a similarity graph), or certain matrices derived from it, in order to detect the communities of the associated graph. 
In particular, spectral clustering algorithms utilize the eigenvalues and eigenvectors of various versions of the Laplacian of the similarity matrix of the graph, to perform dimensionality reduction of the data space in fewer dimensions, before performing the clustering. On this reduced dimension matrix, a standard clustering method is then applied; typically the \textit{k-means} clustering \citep{Macqueen1967}. 
The spectral clustering algorithms' main advantage is that clusters are not assumed to be of any specific shape or distribution, in contrast to e.g. a simple k-means algorithm. This means that the spectral clustering algorithms can perform well with a wide variety of shapes of data.
On the other hand, the main disadvantage of spectral graph clustering algorithms is that generally they require the number of clusters as an input parameter (although there is a heuristic applied for determining this number).

Spectral clustering algorithms may present some difficulties in identifying communities and distinguishing between clusters in a graph, when dealing with datasets with very close (similar) eigenvalues. Therefore, a better approach is to normalize the data before extracting the eigenvalues and eigenvectors from the similarity matrix. As such, we implement the normalized spectral clustering method proposed by \citet{Shi_Malik2000}, and by \citet{Ng2001}.

The algorithm of \citet{Shi_Malik2000} involves the following steps: {\em (1)} the construction of the similarity or {\em affinity} matrix using the Gaussian similarity function: \\ $A_{ij} = exp(- \norm{s_i - s_j}^2 / 2 \sigma^2)$ \\ where    $\sigma$ is a scale parameter (see section \ref{Methodology_Section} for more details), {\em (2)} the computation of an unnormalized Laplacian matrix, {\em (3)} the extraction of its generalized eigenvectors and {\em (4)} finally, the application of $k$-means in this eigenvector space for the detection of the clusters. The algorithm of \citet{Ng2001} follows a similar process, but implements a normalization of the Laplacian matrix eigenvectors before applying a $k$-means clustering algorithm to extract the clusters, in an attempt to improve the accuracy and computational efficiency of the algorithm.
Our study focuses on the second kind of graph clustering algorithms, since the approach allows a more concrete analysis of the quality of the solution. For this reason, the spectral graph clustering algorithms explored here are described in more detail in the methodology section (\ref{Methodology_Section}).

\section{Methodology}\label{Methodology_Section}

In this section we describe the research methodology employed for this work. The first step of the methodology is the construction of a suitable {\em similarity} graph based on galaxy SEDs (spectra). This is obtained by a careful choice of a similarity function (i.e. the Gaussian Kernel) as well as a sparsification method to obtain a sparse graph that will allow for an easier extraction of the graph clusters.

The second step of the methodology is a comparative and careful application of graph clustering algorithms on the resulting similarity graph for the detection of communities (clusters) of galaxies with similar SEDs. The third step of our approach is to exploit the visualization provided by the graph modeling in order to extract possible hidden relations between various physical properties of galaxies (i.e. SEDs with other physical properties, such as the presence of polycyclic aromatic hydrocarbon (PAH) molecules and silicate dust grains) but also to suggest possible evolutionary processes for the power sources of ULIRGs. Additionally, the exploitation of diverse graph clustering algorithms serves as  a verification tool for the algorithms employed, which also provides multi-scale clustering of the galaxies. The last and perhaps most important step of our approach is the interpretation of our results, obtained by comparing them with results from previous works and the established theoretical framework of ULIRGs and their evolution.

For the graph visualization and implementation of network analysis, we utilize several Python packages, such as the \textit{scikit-learn} and \textit{igraph} packages.

\subsection{Construction of the Similarity Graph}\label{Similarity_Section}

We recall that the data set under investigation consists of $139$ (denoted be $N$) ULIRGs and quasars. For each of these galaxies, its mid-IR spectra (of $D$ values) is known, resulting to an $N \times D$ matrix, which will be denoted by $\mathbf{X}$. Additionally, each galaxy's spectrum is characterized by the presence of distinct PAH emission features (at $6.2 \mu m$ and $11.2 \mu m$) and a deep Silicate absorption/emission feature at $9.7 \mu m$.

From  the $N \times D$ matrix  $ \mathbf{X} $ containing galaxy SEDs, we construct a corresponding {\em similarity matrix} utilizing a {\em similarity function} (see Section \ref{GraphClusteringSection}). One of the most commonly used functions - especially for the case of spectral graph clustering - is the {\em Gaussian Radial Basis Function (RBF) kernel}, also known as {\em Gaussian kernel}, which is defined as:
  \begin{equation}
      s(\mathbf{x}_i,\mathbf{x}_j)=\exp{(-\gamma||\mathbf{x}_i-\mathbf{x}_j||^2)},
  \end{equation}
where $||\mathbf{x}_i-\mathbf{x}_j||$ is the Euclidean distance between the vectors $\mathbf{x}_i,\mathbf{x}_j \in \mathbb{R}^D$, and $\gamma = 1/2{\sigma}^2$, where $\sigma$ is the corresponding standard deviation of the Gaussian distribution. In this way, a {\em similarity matrix} $\mathbf{S}$ can be constructed, where $S_{ij} = s(\mathbf{x}_i,\mathbf{x}_j)$. We also note that, by definition, the Gaussian Kernel is a similarity function (not a dissimilarity one): The smallest the difference between $\mathbf{x}_i,\mathbf{x}_j$, the largest is their corresponding $s(\mathbf{x}_i,\mathbf{x}_j)$ value. 
Additionally, we note that the Gaussian kernel function provides a non-linear representation of the data set, through the corresponding similarity distances between them, capturing in this way their non-linear relationships. For the aforementioned reasons, we choose to implement the Gaussian Kernel function for the construction of the similarity matrix and the corresponding graph. 
  
We remark that there exist various suggestions for choosing the parameter $\sigma$. A common and natural choice (e.g. \citet{Veenstra_2016}) is to consider $\sigma$ to be the standard deviation of the sample of the $\frac{N}{2}$ distances $||\mathbf{x}_i-\mathbf{x}_j||$. In this work we utilize this way for choosing the value of $\sigma$. We have tested several other choices for $\sigma$, but have chosen this approach since it has been shown to obtain more natural results for the particular data set under investigation.

From the similarity matrix we can now compute the {\em similarity graph} (and its corresponding adjacency matrix) $G = (V, E, w)$ as follows: $V$ corresponds to the $N$ galaxies under investigation and for each $ i, j \in V $, $ w(i,j) = S _{i,j}$. Note that the resulting graph is a weighted but almost complete graph (see \ref{GraphTheorySection}). Such a dense graph makes both the visualization of the similarities of galaxies hard to distinguish and also makes the clustering analysis difficult \citep{vonluxburg2007tutorial}. A common way to overcome this difficulty, used both for graph clustering as well as more generally for various graph-theoretic analyses, is the {\em graph sparsification}. Through this process, the lightest edges among nodes (corresponding to less significant similarities among galaxies) are removed from the graph. The resulting sparse graph contains the most significant edges among pairs of nodes, representing the most important relations between nodes of the network. In the case of similarity graphs the edges with the largest weights remain while the lightest edges are removed from the graph. Two of the most common sparsification methods used for graph clustering are to apply either a global or a local criterion for choosing which edges are removed: (1) The $\epsilon$-neighbourhood method \citep{vonluxburg2007tutorial} applies a {\em global} threshold value for the whole graph for the decision of which edge to be removed, keeping only edges of weight greater than the threshold. (2) On the other hand, in the $k$-nearest neighbour ($k$-nn) method \citep{vonluxburg2007tutorial}, for each node of the graph, we keep only the   $k$ largest  weighted edges incident to it. In this work, we utilize the $k$-nn method for the graph sparsification, with a careful choice for the value of $k$, as explained in Section \ref{ResultsSection}. The resulting sparse, weighted similarity graph (obtained by applying the Gaussian Kernel on the galaxy SEDs), after the sparsification process is called {\em SED similarity} graph and is denoted as $G(V, E, \mathbf{W} )$ or simply as $G$.

After the graph construction and sparsification of the SED similarity graph, we first apply some basic network analysis on the resulting graph, in order to achieve a better understanding of the data set under investigation from a network modeling point of view, by computing important properties of the network, such as node degree and node centrality (see Section \ref{GraphTheorySection}). 
Then, we proceed to the main step of our graph theoretic analysis, which is the application of graph clustering algorithms, on the obtained similarity graph, in order to detect distinct communities of galaxies with similar spectra. This step is explained in detail in the following section.

\subsection{Application of Modularity Based and Spectral Graph Clustering Algorithms}\label{Graph Clustering Algorithms Section:ModularityBased}

The next step of our analysis is the exploration of different graph clustering algorithms on the resulting SED similarity graph. Since graph clustering algorithms in graphs group together highly connected nodes, these groupings   correspond to  galaxies of similar SEDs, which may correspond to different stages of galaxy evolution and could also reveal hidden relations between various physical properties of galaxies (i.e. a comparison between PAH emission and silicate absorption features). We explore two kinds of graph clustering algorithms: modularity-based ones (applied directly on the adjacency matrix) and spectral graph clustering algorithms which apply $k$-means clustering on the eigenvectors of the Laplacian of the similarity graph. Both methods have been described in Section \ref{GraphTheorySection}.

\subsubsection{Applying Modularity-based Clustering Algorithms}\label{Graph Clustering Algorithms Section:SpectralBased}

We explore the application of two modularity-based algorithms on the SED similarity graph: The Leiden algorithm of \citet{Leiden} and the Walktrap algorithm of \citet{Pascal2006} (see Section \ref{GraphClusteringSection}). The Leiden algorithm returns the clusters for the level of hierarchy that achieves maximal modularity. The Walktrap algorithm also allows for the possibility to  return the clusters obtained for any level of hierarchy. Additionally, the Walktrap algorithm works also by utilizing conductance as a criterion for the  merging of the clusters, instead of modularity score. This is quite useful for the choise of the level of hierarchy which results in clusters that achieve both optimal conductance and optimal modularity (see Section \ref{ResultsSection} for more details).

\subsubsection{Applying Spectral Graph Clustering Algorithms}
 
Additionally, we explore spectral graph algorithms on the SED similarity graph obtained. As described in Section \ref{GraphClusteringSection}, this type of algorithms apply a $k$-means \citep{Macqueen1967}) on the matrix obtained by the $k$ most significant eigenvectors of the Laplacian of the similarity graph under investigation.
  
In particular, we utilize the normalized spectral clustering method proposed by \citet{Shi_Malik2000}. This method involves the construction of an \textit{affinity matrix}, which corresponds to the similarity matrix obtained by the application of the Gaussian Kernel similarity function in the raw data, as we explained in section \ref{Similarity_Section}. 

This clustering method also involves the computation of an unnormalized Laplacian matrix and the extraction of its generalized eigenvectors, which represent the detected number of clusters of the graph. 
The entire process, as implemented in this work is explained in \ref{Appendix_section_ShiMalik}.

As explained in Section \ref{Graph Clustering Algorithms Section:ModularityBased}, spectral graph clustering algorithms require the number of clusters $r$ to be given \textit{a priori} as part of the input. However, there are several related heuristics that compute this value. The most popular method, which we utilize in this work, is the \textit{eigengap} heuristic \citep{vonluxburg2007tutorial}. According to this method, $r$ is the minimum value for which the difference (gap) between the consecutive Laplacian eigenvalues of the Laplacian of the graph $G$,  ${\lambda}_{r+1}-{\lambda}_{r}$ becomes significantly larger (see Section \ref{Appendix_section_ShiMalik} for more details). \par

This is justified by perturbation arguments from the ideal case of $r$ distinct connected components, where clearly ${\lambda}_1 = \cdots =  {\lambda}_r = 0$ and ${\lambda}_{r+1} >0$. \par

The second normalized spectral clustering method we apply is the one proposed by \citet{Ng2001}. The proposed algorithm  differs from the algorithm of \citet{Shi_Malik2000} mainly in two points: {\em (i)} In step (2), it computes the \textit{normalized Laplacian matrix} instead of the unnormalized Laplacian, which is defined as: 
$\mathbf{L}^{norm}  = \mathbf{D} ^{-1/2}     \mathbf{L} \;  \mathbf{D}^ {-1/2}$ and {\em (ii)} Adds an additional step between steps (3) and (4), where the   rows of matrix $\mathbf{X}$ are re-normalized in order to have unit length, obtaining matrix $ \mathbf{Y}$: with $Y_{ij} = X_{ij} / (\sum_j X_{ij}^2)^{1/2}$. Then, step (4) applies on matrix $\mathbf{Y}$ instead of $\mathbf{X}$.

\citet{Ng2001} note that there are many similarities between spectral clustering methods and \textit{Kernel PCA} (KPCA), another method which can be used for clustering analysis. The authors also point out that the main difference between their method and KPCA is the initial normalization of the affinity matrix, which improves the algorithm's performance.

\section{Results}\label{ResultsSection}

In this section we present the results we obtain following the methodology described in previous sections. First, we present the constructed (sparsified) similarity graph based on galaxy mid-IR spectra. \par

Next, we apply several graph clustering algorithms, as explained in section \ref{GraphClusteringSection} to detect communities of galaxies of similar spectra. Next, we compare in a visualized way the SED communities with other physical properties of the galaxies, such as the PAH emission and the silicate absorption features. The methods explored produce communities that correspond to known stages of ULIRG evolution, suggesting an evolutionary path of their power sources. The placement of the galaxies throughout the generated graph provides a visual representation illustrating the similarities between well-known galaxies, as well as relatively unknown ones, that may be of interest for further investigation. Finally, we present an interpretation of the obtained results, combined with the application of domain knowledge, which can lead to verification and improvements upon previous works, as well as novel suggestions for further understanding galaxy evolution. \par

\subsection{The Similarity Graph and Basic Network Analysis}

\begin{figure*}
\centering
\includegraphics[width=0.5\textwidth]{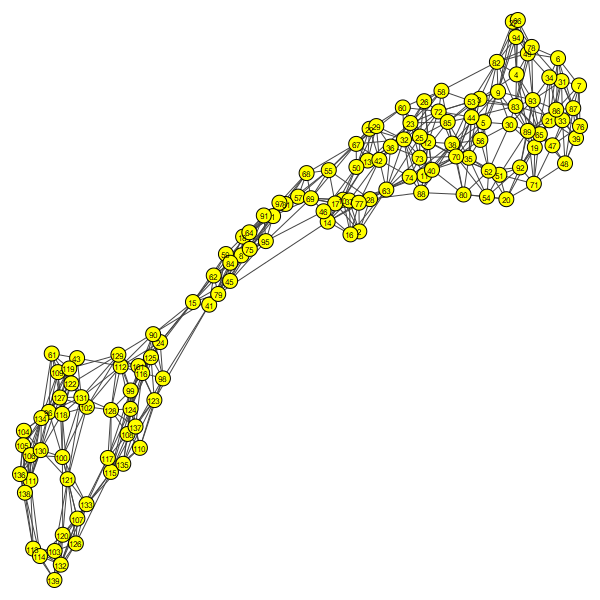}
\caption{A drawing of the resulting (sparsified) similarity graph using the Gaussian Kernel as a similarity function  and the   weighted Fruchterman-Reingold spring layout algorithm of {\sf igraph} for its drawing. The nodes are referenced with numerical labels as presented in Table \ref{labels_table} of \ref{appendixIII}. The network demonstrates a \textit{network diameter = 13}, an \textit{average vertex degree = 9.3}, a \textit{minimum vertex degree = 7}, a \textit{maximum vertex degree = 18} and a \textit{network clustering coefficient = 0.52}.}\label{5PCs_Farrah_PG_Spring_layout}
\end{figure*}

As explained in previous sections, we first use the raw data (i.e. the SEDs of our sample) for the construction of the {\em similarity graph}, using the Gaussian Kernel. In section \ref{Similarity_Section}, we justify our choice for the value of $\sigma$ as the standard deviation based on the Euclidean distances between the vectors $x_i, x_j$. Additionally, for the sparsification of the similarity graph, we are interested in maintaining both connectedness and an adequate representation of the original data. For large graphs, connectedness is asymptotically guaranteed if $k$ is chosen to be of the order of $\log(n)$. For small/medium sized graphs though, as in our case, some exploration is required. In our case and for the $k$-nn method, we check the connectedness of different graphs resulting from increasing values of $k$. The graphs for $k$-nn are connected for $k \geq 3$, but $k=3$ is too sparse. For the purposes of our work we use exclusively the \textit{mutual k-nn} method, in which the k-nearest neighbors are connected via edges which are undirected, meaning that the $k$-nn connection is mutual between the nodes. The graphs for mutual $k$-nn might be disconnected, but for increasing $k$ the size of the largest connected component increases. This growth stabilizes for $k=6$ with $\sim 91\%$ of the nodes being included to this component, while the rest of the nodes are isolated points. Since the mutual $k$-nn graph is always a subgraph of the standard $k$-nn graph we simply choose $k=6$ for the standard $k$-nn graph. \par

A drawing of the resulting sparsified graph (in a two dimensional space) is shown Figure \ref{5PCs_Farrah_PG_Spring_layout}. The nodes in this figure (and all of the subsequent figures in this work) are referenced with numerical labels as presented in Table \ref{labels_table} of \ref{appendixIII}. For the drawing, we use the force-directed layout the Fruchterman-Reingold Algorithm \citep{Fruchterman1991}. Force-directed layout algorithms \citep{Kobourov2012} produce graph drawings (layouts) of as few crossing edges as possible, by assigning repulsive forces between all nodes and attractive forces between connected nodes, so that neighbouring nodes are placed closely in the plane. \par

After the construction of the similarity graph, we first apply some basic network analysis in order to understand the system properties. In particular, we compute the following (global) network properties: ({\em  i}) network diameter, ({\em  ii}) average, min max vertex degrees and  ({\em  iii}) network  clustering coefficient (which measures how well the neighbours of a node are neighbors to each other).  The values of these parameters for the constructed network are shown under Figure \ref{5PCs_Farrah_PG_Spring_layout}. We observe that the network demonstrates the following properties:  
({\em  i}) Network diameter = 13,
    ({\em  ii}) Average vertex degree = 9.3,
    ({\em  iii}) Minimum vertex degree = 7,
    ({\em  vi}) Maximum vertex degree = 18 and
     ({\em  v}) Network clustering coefficient = 0.52.

The network diameter indicates that the maximum number of shortest paths between connected nodes in the graph is $13$. The vertex degree of a graph represents the number of connections to each node. Our graph shows an average degree for each node $= 9.3$, with the minimum and maximum number of connections being $7$ and $18$, respectively. The calculated network clustering coefficient ($0.52$) demonstrates that around half the neighbors are connected to each vertex in the graph (with a value of 0 indicating that no neighbors are connected and a value of 1 indicating that all neighbors are connected to each vertex). \par

To better understand these results, we can compare them to the corresponding metrics produced by a random graph. For this purpose we have calculated the corresponding metrics of a random graph generated based on the \href{https://igraph.org/python/api/latest/igraph._igraph.GraphBase.html#Erdos_Renyi}{Erdos-Renyi} stochastic model. For this implementation we only provide the same number of nodes ($139$) and number of total edges ($834$) as produced during the generation of our similarity graph. This results in a network diameter $= 3$ for the random graph, which is much lower than the network diameter of our similarity graph ($= 13$), suggesting that a random graph would result in a much more connected network, as expected. This showcases that the nodes in our similarity graph are more widely separated, suggesting a more meaningful structure of the network (with similar nodes being placed much closer together compared to dissimilar ones). Furthermore, whereas the minimum and maximum vertex degrees of the random graph ($= 6$ and $= 20$ respectively) showcase similar values to our similarity graph, the average vertex degree of the random graph is $= 12$, which further indicates a more highly connected network compared to our similarity graph, where the average vertex degree is $= 9.3$. Finally, the network clustering coefficient of the Erdos-Renyi random graph is $=0.09$, a much lower value than that of our similarity graph ($= 0.52$) which showcases that the neighboring nodes of the random graph are not as well connected as in our similarity graph. \par

\subsection{Extraction of SED Communities}\label{ExtractionSED_Communities_Section}

We next apply graph clustering on the SED similarity graph, to identify groups of galaxies with similar spectra. In the following figures we present the results obtained by applying several graph clustering algorithms, as described in Section \ref{Graph Clustering Algorithms Section:ModularityBased}. In all of the following graph drawings, different clusters detected are represented with a different color in the corresponding node. \par

\begin{figure*}
\begin{multicols}{2}
\begin{subfigure}{.5\textwidth}
\centering
\includegraphics[width=0.875\textwidth]{5PCs_Farrah_PG_Leiden}
\caption{}\label{Communities_Leiden}
 \end{subfigure}

\begin{subfigure}{.5\textwidth} 
\centering
\includegraphics[width=0.875\textwidth]{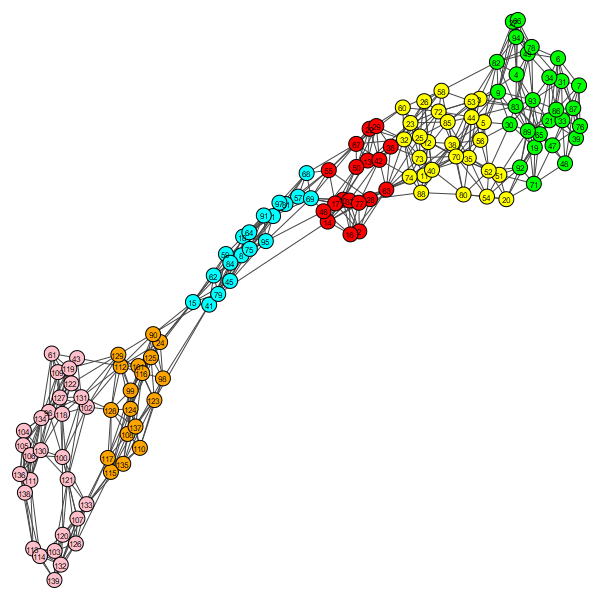}
\caption{}\label{Communities_Walktrap}
\end{subfigure}
\end{multicols} 

\caption{The application of Modularity based algorithms (Leiden and Walktrap) results to the detection of six Communities of very similar quality. (a) Left: Communities detected by applying the Leiden algorithm of \citet{Leiden} on the resulting similarity graph. The algorithm detected $6$ communities  and achieved a modularity score of $0.7$ and conductance of $0.30$. (b) Right: Communities detected by applying the Walktrap algorithm of \citet{Pascal2006} on the similarity graph. After $30$ steps (iterations of hierarchy levels), this algorithm detected $6$ communities, with a modularity score of $0.7$ and a max conductance of $0.21$.}\label{modularity_communities_figure}

\end{figure*}

\begin{figure*}
\begin{multicols}{2}
\begin{subfigure}{.5\textwidth}
\centering
\includegraphics[width=0.875\textwidth]{5PCs_Farrah_PG_Andrew_Ng_graph}
\caption{}\label{Andrew_Ng_communities}
\end{subfigure}

\begin{subfigure}{.5\textwidth}
\centering
\includegraphics[width=0.875\textwidth]{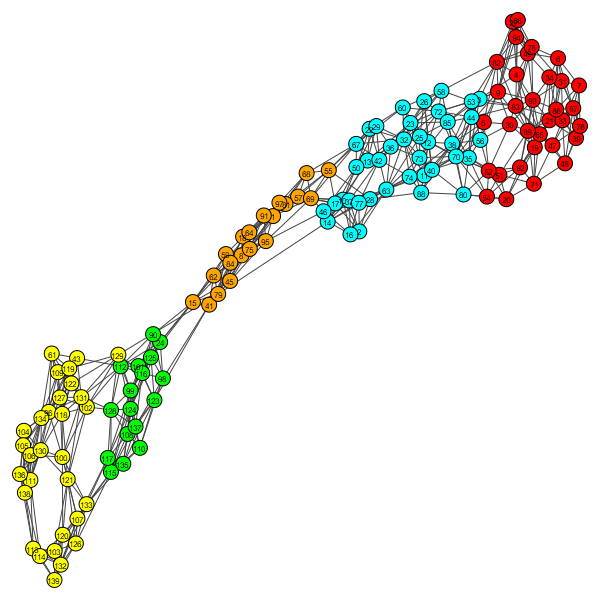}
\caption{}\label{Shi_Malik_communities}
\end{subfigure}

\end{multicols} 

\caption{The application of Spectral graph clustering algorithms (Ng \&  Jordan and Shi \& Malik) which results in the detection of five Communities of very similar quality. (a) Left: Communities detected by the spectral graph clustering algorithm of Andrew Ng \& Jordan \protect\citep{Ng2001}, by setting the parameter of the number of clusters to be returned to $5$. The conductance obtained is $0.05$, with a modularity score = $0.68$. (b) Right: Communities detected  by the spectral graph clustering algorithm of \protect\citet{Shi_Malik2000}, by setting the parameter of the number of clusters to be returned to $5$. The conductance obtained is $0.05$, with a modularity score = $0.68$.}\label{spectral_communities_figure}

\end{figure*}

We begin by presenting the results obtained with the implementation of Modularity based algorithms. Figure \ref{Communities_Leiden} shows the result obtained by applying the Leiden algorithm of \citet{Leiden}. Six communities were detected, with an optimal modularity score of $0.7$ and a conductance score of $0.30$. This result was achieved via the {\sf igraph} implementation of the algorithm, using a  resolution parameter of $1.0$ (the resolution parameter determines the granularity of the clustering; a higher resolution parameter corresponds to a lower number of detected communities). We tested several values for the associated resolution parameter. The chosen value achieves the best results, in terms of both the quality of the communities detected as well as their physical meaning. The second Modularity based algorithm explored is the Walktrap algorithm of \citet{Pascal2006}. The result obtained by the application of this algorithm on our graph is shown in Figure \ref{Communities_Walktrap}. Again, six communities were detected, with a modularity score of $0.7$. Furthermore, the algorithm achieves a conductance score of $0.21$. \par

\begin{figure}[H]
\centering
\includegraphics[width=1\columnwidth]{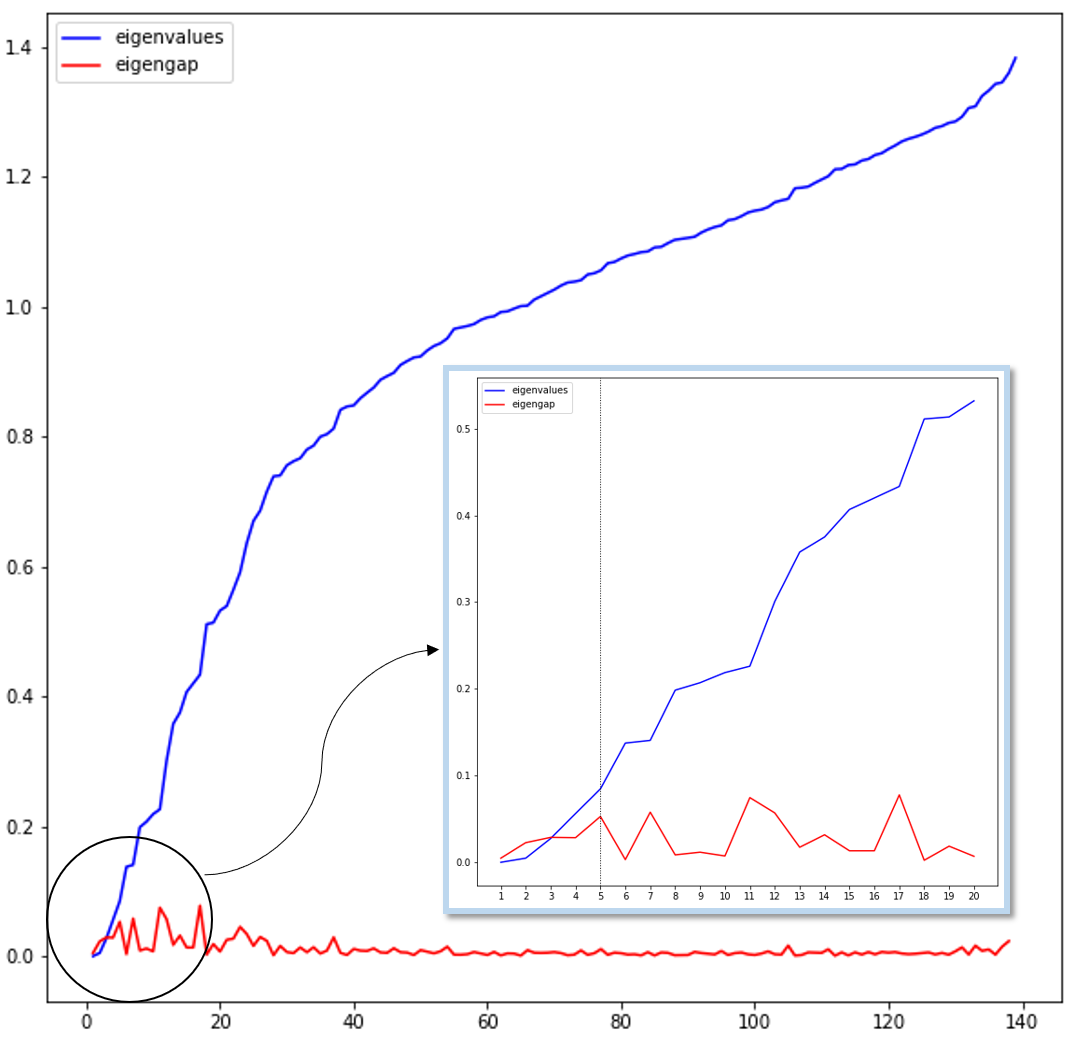}
\caption{The eigenvalues $\lambda_i$ of the generalized eigen-problem ${\mathbf L}u = {\lambda}_i u$ for the unnormalized Laplacian (blue) and the corresponding eigengaps (red). Applying the eigengap heuristic on the smallest $20$  Laplacian eigenvalues, the significant gap is found between ${\lambda}_5$ and ${\lambda}_6$, hence we have $r = 5$ clusters. }\label{LaplacianEigengap}
\end{figure}

Next, we present the results obtained applying spectral graph clustering algorithms, as explained in Section \ref{Graph Clustering Algorithms Section:SpectralBased}. 
Spectral graph clustering algorithms require the number of clusters to be provided as part of the input. Applying the eigengap heuristic shown in Figure \ref{LaplacianEigengap}, the number of clusters $r $ is chosen to be $r=5$. Based on the methodology explained in section \ref{Graph Clustering Algorithms Section:SpectralBased}, in order to determine this value, we determine that the minimum value which combines {\em (i)} the spectral gap heuristics and {\em (ii)} also achieves both the optimal modularity and conductance scores is the value of $5$. Hence we choose this value as input for the spectral algorithms explored in our work. \par

Figure  \ref{Andrew_Ng_communities} shows the five communities detected   by applying the algorithm of \citet{Ng2001} and figure \ref{Shi_Malik_communities} shows the five communities detected by applying the algorithm of \citet{Shi_Malik2000}.
Interestingly, both algorithms achieve similar conductance values of $0.05$, which is much lower than the corresponding values achieved by modularity based algorithms (between $0.2 - 0.3$). \par

\subsubsection{Comparing Communities detected by various algorithms}

In the previous section we show a qualitative comparison of the outputs of the application of several graph clustering algorithms on our dataset, in which all graph clustering algorithms obtain very similar results visually. In this section, we present a quantitative analysis for comparing these outputs. In particular we would like, for each pair of the algorithms tested, to {\em (i)} match the communities detected between the two and {\em (ii)} after the mapping of corresponding communities between the outputs of the two algorithms, compute the percentage of nodes that are common in both of the outputs. \par

\begin{table}[H]
    \centering
    \includegraphics[width=1\columnwidth]{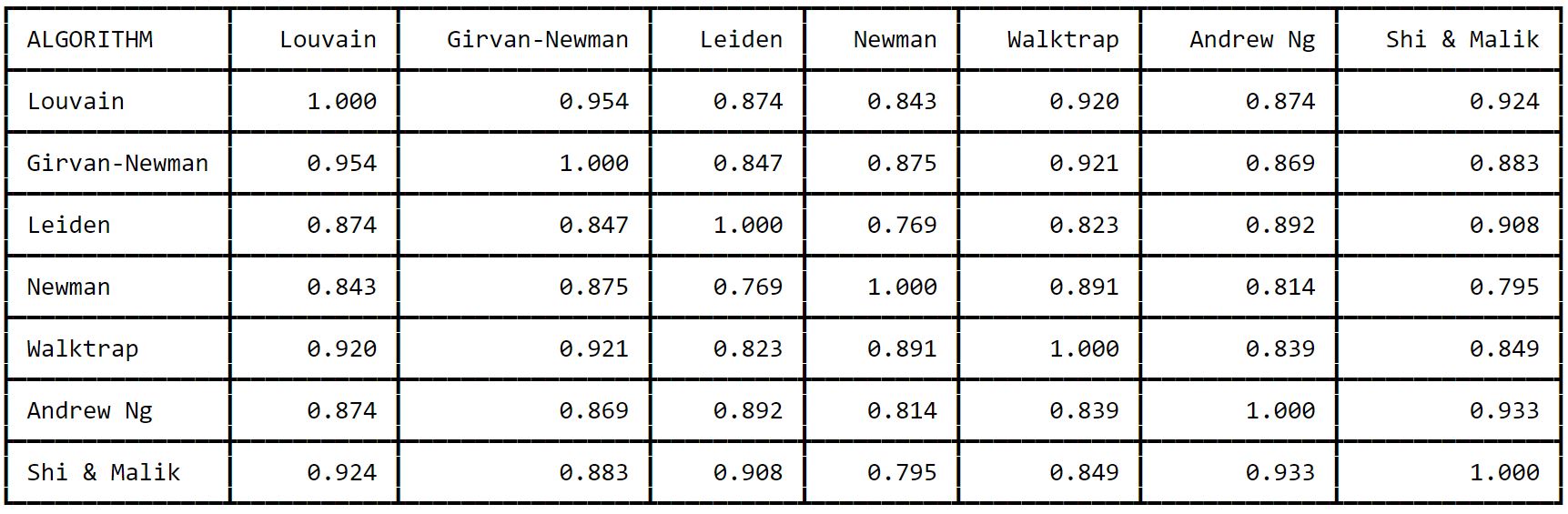}
    \caption{A pairwise comparison of the graph clustering algorithms explored in this work showing the percentage of common nodes between  (corresponding) detected clusters.}
    \label{table:comparing_algorithms_table}
\end{table}

Table \ref{table:comparing_algorithms_table} shows the results of this comparison. We observe a very high percentage of matching of detected clusters between different clustering algorithms, with an average matching percentage between all algorithms $= 86.6\%$. This means that an average of $120$ galaxies of our sample of $139$ galaxies are placed in similar communities under all clustering algorithms considered in this work. \par

The matching percentage between modularity-based clustering algorithms is $= 87.6\%$, with the lowest matching percentage being $= 78.7\%$ (between the Leiden and Newman algorithms). The matching percentage between the two spectral graph clustering algorithms of \citet{Ng2001} and \citet{Shi_Malik2000} is $= 93.3\%$. This concludes that the implementation of the different types of clustering algorithms in our graph results in very similar results of detected clusters, especially considering that the optimal number of clusters for modularity-based algorithms (6 clusters) is different than the optimal number of clusters for spectral graph clustering algorithms (5 clusters). We are therefore highly confident on the consistency and nature of the detected communities based on our methodology. \par

Another interesting observation is the high matching correlation ($93.3\%$) between the clusters obtained by the application of spectral graph clustering algorithms. This means that $130$ of the $139$ galaxies investigated in this work are placed in matching communities under different spectral graph clustering algorithms. \par 

These results serve as a verification of the validity of the graph clustering algorithms investigated, as well as a demonstration on the suitability of graph theory for investigating similar problems in galaxy evolution and astrophysics. \par

It should be noted at this point that the difference in the number of detected clusters between the modularity-based algorithms and the spectral graph clustering algorithms arises due to the difference in the methodology of these two distinct types of clustering algorithms. The modularity-based algorithms consistently detected six clusters, based on the optimization of the modularity score in each implementation. Comparatively, as mentioned previously, the spectral graph clustering algorithms take the number of detected clusters as a parameter for their implementation. In this case, we use a purely mathematical justification for determining the appropriate number of clusters, based on the Laplacian eigengap heuristic (\citet{vonluxburg2007tutorial}) method presented in Figure \ref{LaplacianEigengap}. Using this mathematical justification for the selection of the number of clusters, we observe an improved clustering performance by the spectral graph clustering algorithms compared to the modularity-based algorithms, as is evident by the resulting maximum conductance scores of each algorithm, which is much lower in the case of the former ($0.05$) compared to the latter ($0.2-0.3$). This result represents a quantitative demonstration of the improved clustering performance for the case of the spectral graph clustering algorithms on our similarity graph. \par

Based on these results, we have chosen to further investigate the physical features of our sample in following sections using the communities detected by the implementation of the \citet{Shi_Malik2000} algorithm for the purposes of this work. The communities detected from the implementation of the clustering algorithms considered in this work are presented in \ref{Appendix_Clustering_Graphs}. \par

\subsection{Astrophysical Interpretation of Communities}
  
In this section, the graph drawing of the network modeling of the set of galaxies under investigation is explored, in order to first relate the communities detected with physical properties of the galaxies (i.e. PAH emission and silicate absorption features) and secondly to relate and compare the graph theoretical classification obtained in this work with other classification methods, and in particular the well known galaxy classification \textit{fork} diagram introduced by \citet{Spoon2007}.

The graphs in Figures \ref{Andrew_Ng_communities} and \ref{Shi_Malik_communities} present the clusters detected using the Spectral Graph Clustering algorithms of \citet{Ng2001} and \citet{Shi_Malik2000}, respectively, using a preset number of clusters $= 5$. As mentioned previously, the implementation of these algorithms result in the detection of similar communities ($93.3\%$ match). Two communities of mostly starburst-dominated ULIRGs and two communities of mostly AGN-dominated quasars are identified, as well as one community of ULIRGs dominated by both starburst and AGN emission, which can be interpreted as a transitional phase between the distinct starburst and AGN-dominated phases. \par

\begin{figure*}
\begin{multicols}{2}

\begin{subfigure}{1.\textwidth}
\centering
\includegraphics[width=1.0\columnwidth]{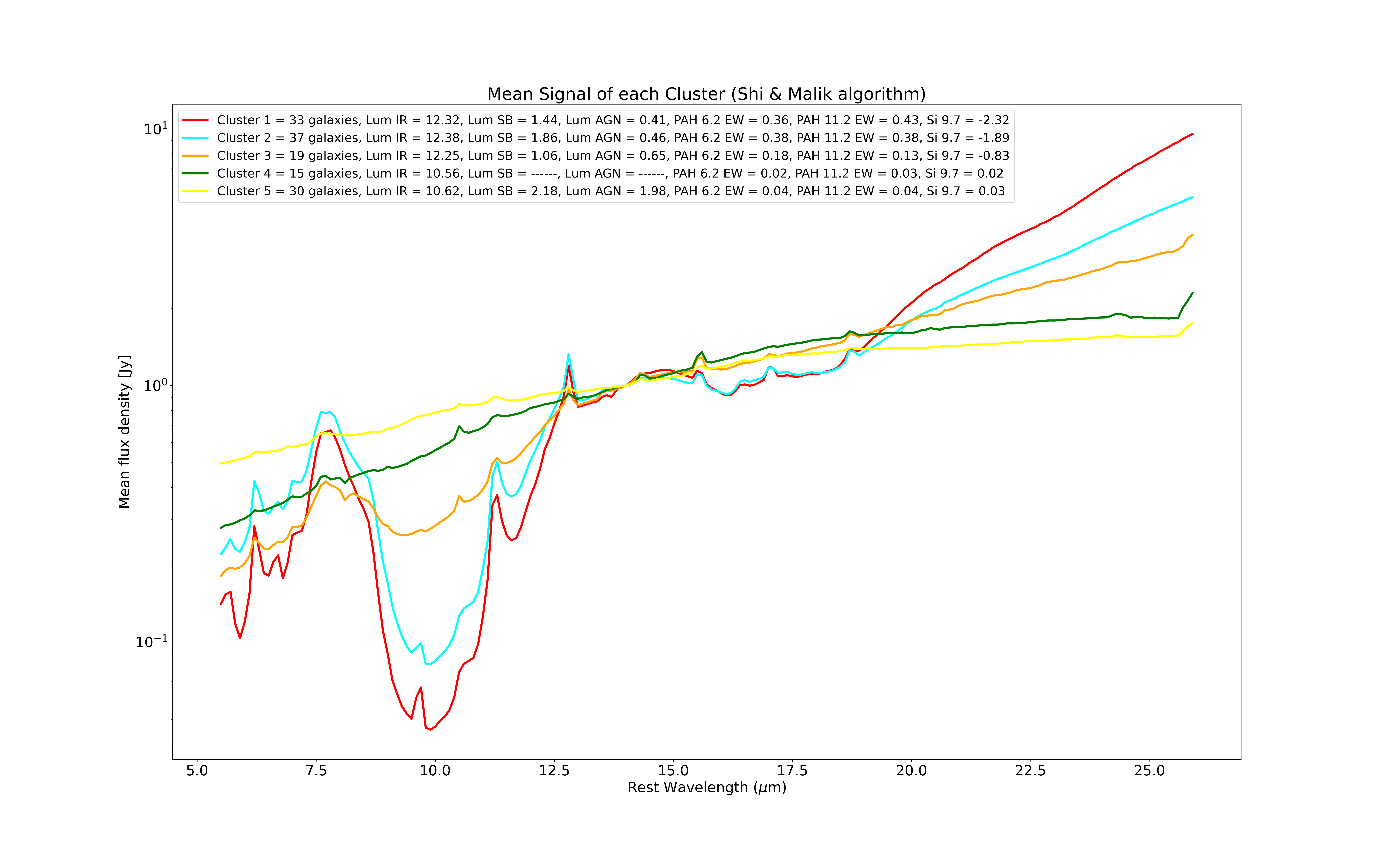}
\end{subfigure}

\end{multicols} 

\caption{Mean mid-IR signals of each cluster detected via the \protect\citet{Shi_Malik2000} algorithm. Each signal represents the mean flux density calculated based on the galaxies of each cluster, normalized at $14 \mu m$. The mean IR luminosity (Lum IR) in units $log_{10}(L_{IR}/L_{\odot})$, the mean starburst luminosity (Lum SB) in units of $10^{12} L_{\odot}$ taken from \citet{Efstathiou2022}, the mean AGN luminosity (Lum AGN) in units of $10^{12} L_{\odot}$ taken from \citet{Efstathiou2022}, the mean PAH equivalent width values at $6.2\mu m$ (PAH 6.2 EW) and $11.2\mu m$ (PAH 11.2 EW), as well as the mean silicate absorption strength at $9.7\mu m$ (Si 9.7) of each cluster are shown in the legend on the top-left. Red and cyan-colored signals display mostly starburst-dominated ULIRG characteristics, the orange-colored signal is displays a mixture of starburst-dominated and AGN-dominated (mostly Seyfert-2) galaxies, and the green and yellow-colored signals represent AGN-dominated quasars (Seyfert-1) galaxies. The extinction of PAH emission and dust absorption features is apparent in the mean signals of AGN-dominated galaxies (green and yellow galaxies).}\label{Mean_Signals_graph}

\end{figure*}

In Figure \ref{Mean_Signals_graph} we plot the calculated mean interpolated signals (flux density over rest wavelength) of each cluster identified by the \citet{Shi_Malik2000} algorithm. The signals were normalized at $14 \mu m$, in order to directly compare the signals' features. Additionally, we display certain features of the mid-IR spectra of ULIRGs for each cluster in the legend of Figure \ref{Mean_Signals_graph}, such as the mean IR luminosity (scaled as $log_{10}(L_{IR}/L_{\odot})$), the mean starburst luminosity (in units of $10^{12} L_{\odot}$) obtained from \citet{Efstathiou2022}, the mean AGN luminosity (in units of $10^{12} L_{\odot}$) obtained from \citet{Efstathiou2022}, the calculated mean PAH equivalent width values at $6.2\mu m$ and $11.2\mu m$, as well as the estimated mean silicate absorption strength at $9.7\mu m$ (Si 9.7). These mean signals, as well as the aforementioned mean features of each cluster suggest that the clustering algorithms managed to recover the known optical spectral types of the galaxies, namely H$_{II}$ galaxies, the low-ionization nuclear emission-line regions (LINERs), as well as the Seyfert-2 galaxies and a combination of Seyfert-1 galaxies and luminous quasars \citep[e.g.][]{Lisa_2006,imanishi07,Yuan_2010}. More specifically, the galaxies of the \textit{red} and \textit{cyan} clusters display starburst characteristics, such as prominent PAH emission and silicate dust absorption signals, in line with the emission signals of \textit{LINER} and \textit{$H_{II}$} galaxies. The clusters on the other end of the graph showcase AGN-luminous quasars properties: The \textit{green} and \textit{yellow} clusters display extinguished PAH emission and silicate dust absorption signals, in line with \textit{Type-I Seyfert} luminous quasars. An intermediate phase of both starburst and AGN-dominated galaxies is identified in the \textit{orange} cluster. The mean signal of the orange cluster also indicates the inclusion of many \textit{Type-II Seyfert} galaxies: AGN-dominated galaxies that are observed with a steep viewing angle (edge-on), where the SMBH in their core is heavily shrouded by a dusty torus (hence the presence of a moderate silicate dust absorption feature). This suggests a transition between the evolutionary phases of the power sources of the host galaxies, from the top-right to the bottom-left of our graph (red-cyan-orange-green-yellow clusters). This interpretation of our results is in agreement with \citet{Hurley1}. \par

Further examination on the nature of the galaxies in each cluster, can be made by investigating the nuclear separation of the host nuclei of each galaxy, shown in \ref{Nuclear_separation_graph}, as well as the optical spectral type of each galaxy, as shown in \ref{Optical_Spectral_Type_graph}. The sample's nuclear separation distances in \ref{Nuclear_separation_graph} are presented in colors (for the available data), similar to \citet{Farrah2009} as: Cyan: $>12$ kpc, Green: $6-12$ kpc, Yellow: $0.1-6$ kpc and Red: $<0.1$ kpc (single nucleus). Noticeably, the galaxies displaying the highest nuclear separations (cyan and green) are concentrated on the top-right of our graph, matching the cyan and red colored clusters identified by the \citet{Shi_Malik2000} spectral algorithm. Inversely, the galaxies with a single nucleus (red) are mostly concentrated in the middle and bottom-left of our graph, where the AGN-dominated clusters are identified by the algorithm. The low-intermediate nuclear separation galaxies (yellow) are spread throughout the graph.

The optical spectral types of the galaxies in our sample (where available) are presented in color in \ref{Optical_Spectral_Type_graph}, similar to \citet{Farrah2009}. Specifically, cyan galaxies represent $H_{II}$, green galaxies are LINERs, while Seyfert-2 galaxies are shown in yellow and Seyfert-1/quasars are shown in red color. As expected, the starburst-dominated $H_{II}$ and LINER galaxies are concentrated in the cyan and red clusters. Seyfert-2 galaxies are mostly present in the orange cluster and Seyfert-1/quasars are mostly seen in the yellow and green clusters, with some also present in the orange cluster. This is also another validation and verification of the detected clusters from our implementation of the \citet{Shi_Malik2000} spectral algorithm.

Therefore, the clusters detected by the implementation of the \citet{Shi_Malik2000} algorithm can be interpreted as:
\begin{itemize}
    \item \textit{Red} cluster = Starburst dominated ULIRGs, mostly LINERs, with high PAH emission and silicate dust absorption features.
    \item \textit{Cyan} cluster = Starburst dominated ULIRGs, mostly $H_{II}$ galaxies, with high-to-moderate PAH emission and silicate dust absorption features.
    \item \textit{Orange} cluster = Both starburst and AGN dominated ULIRGs, with somewhat diminished PAH emission and a moderate silicate dust absorption feature. This is caused by the presence of mostly Type-II Seyfert galaxies, as well as some Type-I Seyfert/quasars, in this cluster.
    \item \textit{Green and yellow} clusters = AGN dominated luminous quasars - Type-I Seyfert galaxies, with a single nucleus, showing completely extinguished PAH emission and silicate dust absorption features, as the AGN is the main power source.
\end{itemize}

\vspace{2mm}

\begin{figure*}
\begin{multicols}{3}
\begin{subfigure}{.3\textwidth}
\centering
\includegraphics[scale=0.275]{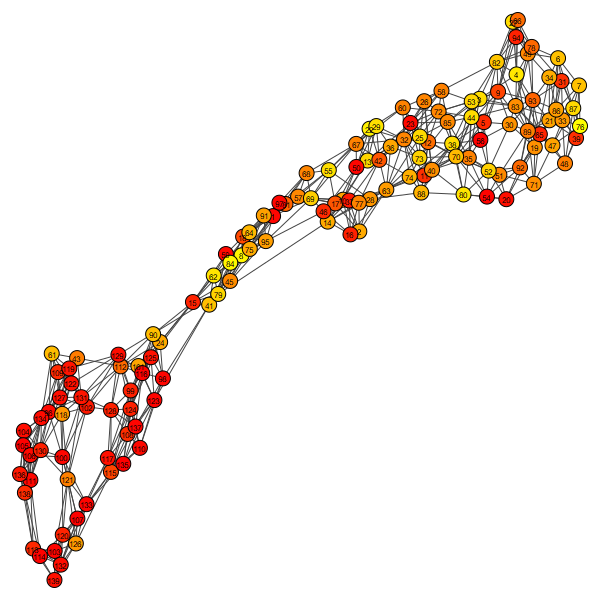}
\caption{ \; }\label{Colormap_PAH_6.2}
\end{subfigure}

\begin{subfigure}{.3\textwidth}
\centering
\includegraphics[scale=0.275]{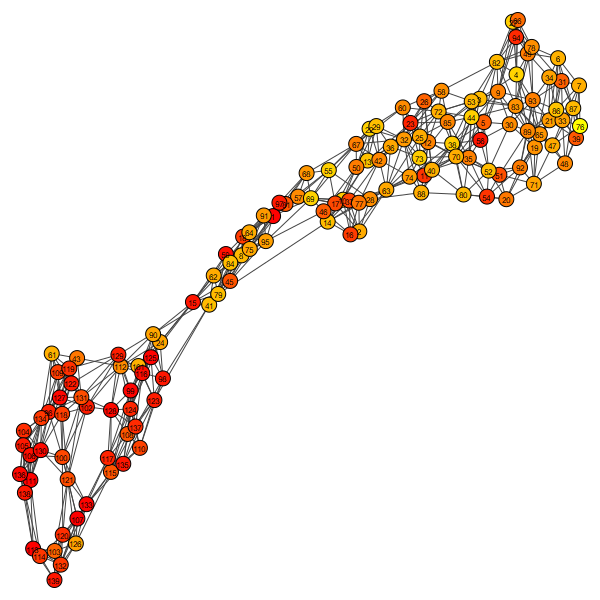}
\caption{ \; }\label{Colormap_PAH_11.2}
\end{subfigure}

\begin{subfigure}{.3\textwidth}
\centering
\includegraphics[scale=0.275]{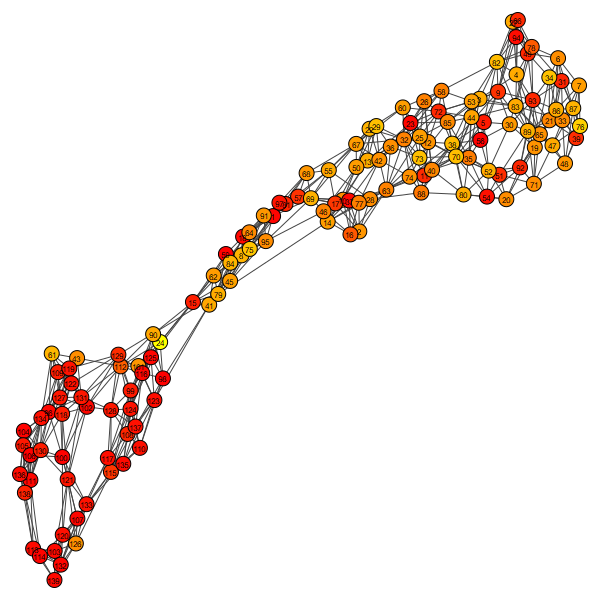}
\caption{ \; }\label{Colormap_Si}
\end{subfigure}
\end{multicols} 

\caption{Colormaps of our graph based on (a) the $6.2 \mu m$ PAH equivalent width emission, (b) the $11.2 \mu m$ PAH equivalent width emission and (c) the Silicate dust absorption of galaxies. Yellow colors denote higher PAH equivalent widths and deeper silicate absorption, whereas red colors denote extinct values for PAH equivalent widths and silicate absorption. Orange nodes correspond to moderate values of PAH equivalent widths and silicate absorption.}\label{Colormaps_PAH_Si}

\end{figure*}

In Figures \ref{Colormap_PAH_6.2}, \ref{Colormap_PAH_11.2} and \ref{Colormap_Si} we present colormaps based on the calculated $6.2 \mu m$, $11.2 \mu m$ PAH equivalent widths and the $9.7 \mu m$ silicate dust absorption feature of our galaxy samples. ULIRGs with high PAH equivalent width values and deep silicate absorption are colored yellow, whereas ULIRGs with extinct PAH and silicate features are colored red. ULIRGs showcasing intermediate (moderate) values of PAH emission and silicate absorption are colored orange. As expected, the colormaps showcase a trend of higher values of PAH emission and silicate dust absorption features for starburst galaxies on the top-right part of the graph and lower values for AGN-dominated luminous quasars towards the bottom-left of the graph. This result is in agreement with the theoretical model proposed by the merger scenario for ULIRGs: As these galaxies interact in the pre-merger stage, they display high PAH emission due to the increase in star-forming activity and the presence of more star-forming regions becomes more prominent (Figure \ref{Colormap_PAH_6.2}, Figure \ref{Colormap_PAH_11.2}). After the coalescence phase, the galaxies merge and their SMBHs become active. This results in the AGN IR emission becoming more dominant and the extinction of the PAH emission features. \par

Similarly, the silicate dust absorption feature in Figure \ref{Colormap_Si}, correlated with the presence of a dust obscured AGN (buried AGN), becomes less distinguishable and almost extinct in the post-merger phase (bottom left of the graph). The galaxies in the intermediate phase (middle of the graph) still showcase a silicate dust absorption feature, which would suggest the presence of an AGN obscured by a dusty torus at this stage. \par

\subsubsection{Comparison of detected communities with physical properties}\label{Comparing_SED_Communities_Section}

An open question in galaxy classification is whether the communities detected using the galaxy spectra are directly correlated to their physical properties, particularly their PAH emission features as well as their silicate absorption/emission feature. Graph theory allows for this multi-variable comparison of galaxies' physical properties in a visual way, by drawing a colored graph, where the color of the node indicates its assigned community and at the same time assign nodes sizes analogous to one of the physical properties under investigation (i.e. PAH emission and Silicate absorption/emission features). \par 

\begin{figure*}
\begin{multicols}{2}
\begin{subfigure}{.4\textwidth}
\centering
\includegraphics[scale=0.4]{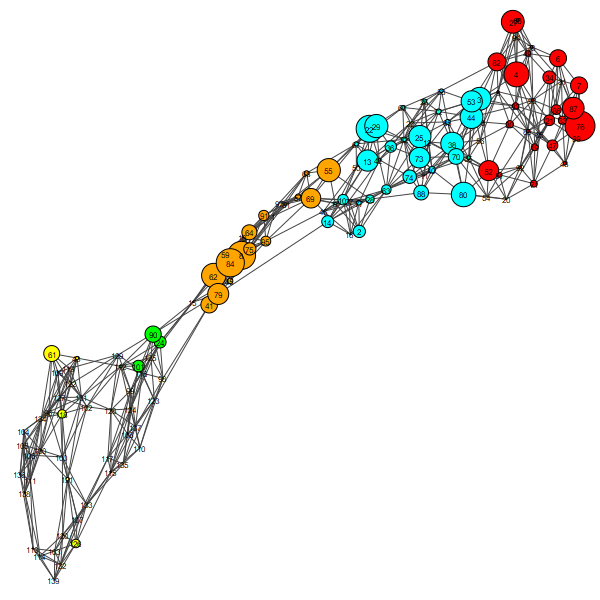}
\caption{ \; }\label{ClustersGraphGaussianPCA_PAH}
\end{subfigure}

\begin{subfigure}{.4\textwidth}
\centering
\includegraphics[scale=0.4]{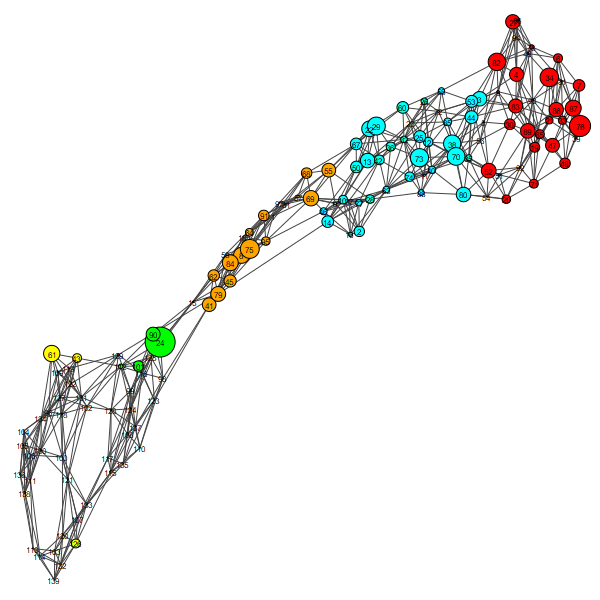}
\caption{ \; }\label{ClustersGraphGaussianPCA_Si}
\end{subfigure}
\end{multicols} 

\caption{Communities detected using the spectral clustering algorithm of \protect\citet{Shi_Malik2000}, in different colors. The node sizes represent (a) the PAH equivalent width emission feature and (b) the Silicate/emission feature. ULIRGs with high PAH emission equivalent widths and deeper silicate absorption features correspond to larger node sizes, whereas ULIRGs with extinct values for both features correspond to smaller node sizes. The transition from weak to strong PAH and Si features, between PG quasars (orange - cyan clusters) and starburst galaxies (green - yellow clusters), is clearly observable.}\label{Shi_Malik_PAH_Si}

\end{figure*}

Figure \ref{ClustersGraphGaussianPCA_PAH} shows our graph with node colors indicating their corresponding community detected using the \citet{Shi_Malik2000} spectral graph clustering algorithm and node sizes analogous to their PAH equivalent width emission feature at $6.2 \mu m$. In Figure \ref{ClustersGraphGaussianPCA_Si} we present the same graph, colored based on the \citet{Shi_Malik2000} algorithm detected clusters and the node sizes corresponding to their silicate absorption feature. In both figures, nodes with larger sizes represent ULIRGs with high PAH equivalent width values and deeper silicate absorption features, respectively. Conversely, nodes with smaller sizes represent ULIRGs with extinct PAH equivalent width and silicate absorption features. The distinction of different values of these features between these communities is immediately apparent. The red and cyan-colored clusters on one end of the graph display higher overall PAH emission and silicate absorption features, compared to the yellow and green-colored clusters at the other end of the graph. These features indicate that starburst activity is more evident in the red and cyan-colored clusters, whereas they become mostly extinct in the yellow and green-colored clusters, suggesting the presence of AGN activity in these galaxies. \par

By viewing figures \ref{ClustersGraphGaussianPCA_PAH} and \ref{ClustersGraphGaussianPCA_Si} in conjunction with the fork diagrams in figures \ref{ClustersGraphGaussianPCA_PAH_6.2} and \ref{ClustersGraphGaussianPCA_PAH_11.2} we can gauge the success of the implementation of the \citet{Shi_Malik2000} algorithm and our overall methodology in classifying galaxies based on the physical properties of their mid-IR emission. The clusters detected clearly correspond to different evolutionary stages of the merger scenario of ULIRGs. Therefore, our graph could be interpreted as a quasi-linear progression in the evolution of ULIRGs, from the starburst-dominated pre-merger stage (red and blue clusters) towards the AGN-dominated post-merger stage (green and yellow clusters). \par

\subsection{Comparison with Related Works}

\subsubsection{The fork diagram}\label{fork_diagrams}

\begin{figure*}
\begin{multicols}{2}
\begin{subfigure}{.4\textwidth}
\centering
\includegraphics[scale=0.32]{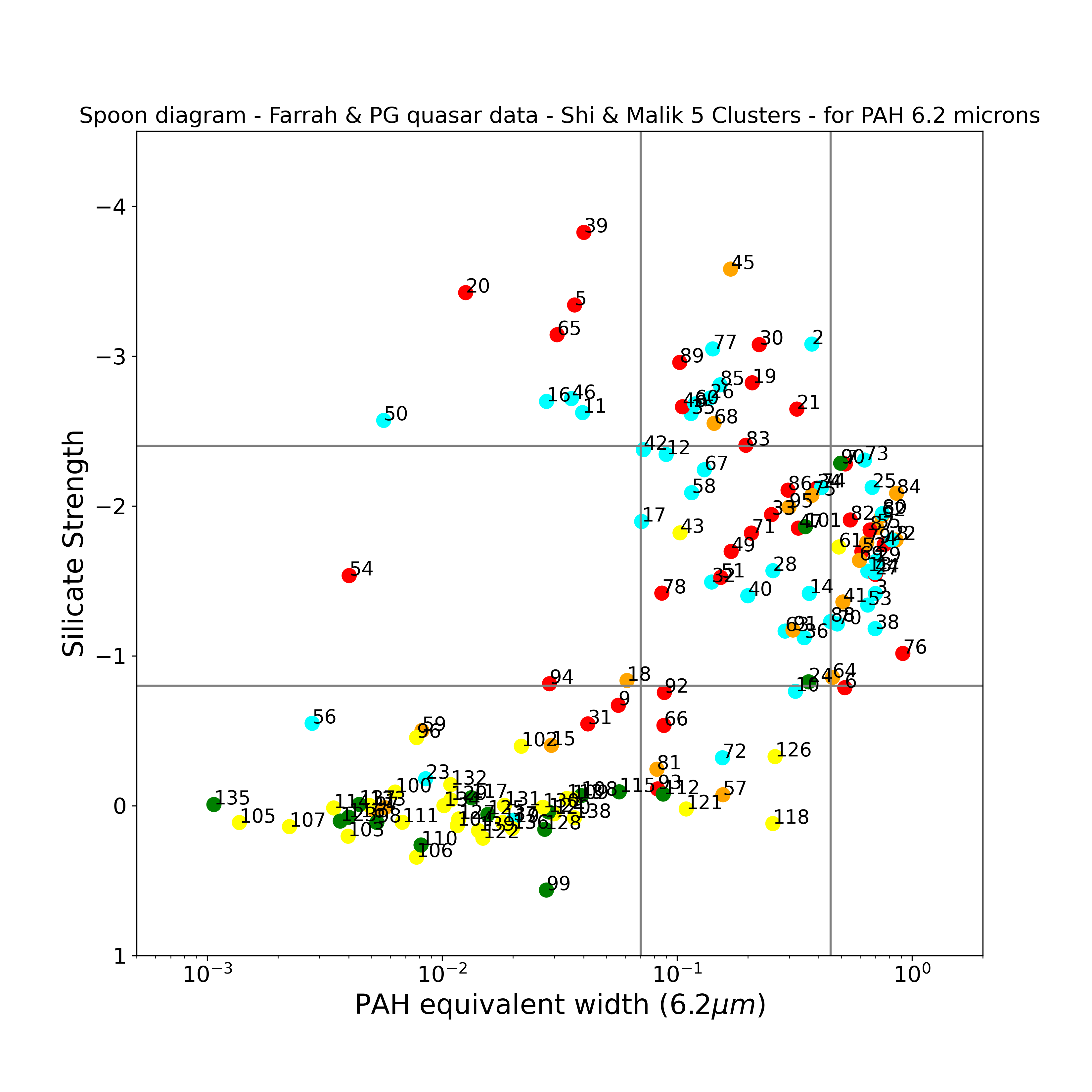}
\caption{ \; }\label{ClustersGraphGaussianPCA_PAH_6.2}
\end{subfigure}

\begin{subfigure}{.4\textwidth}
\centering
\includegraphics[scale=0.32]{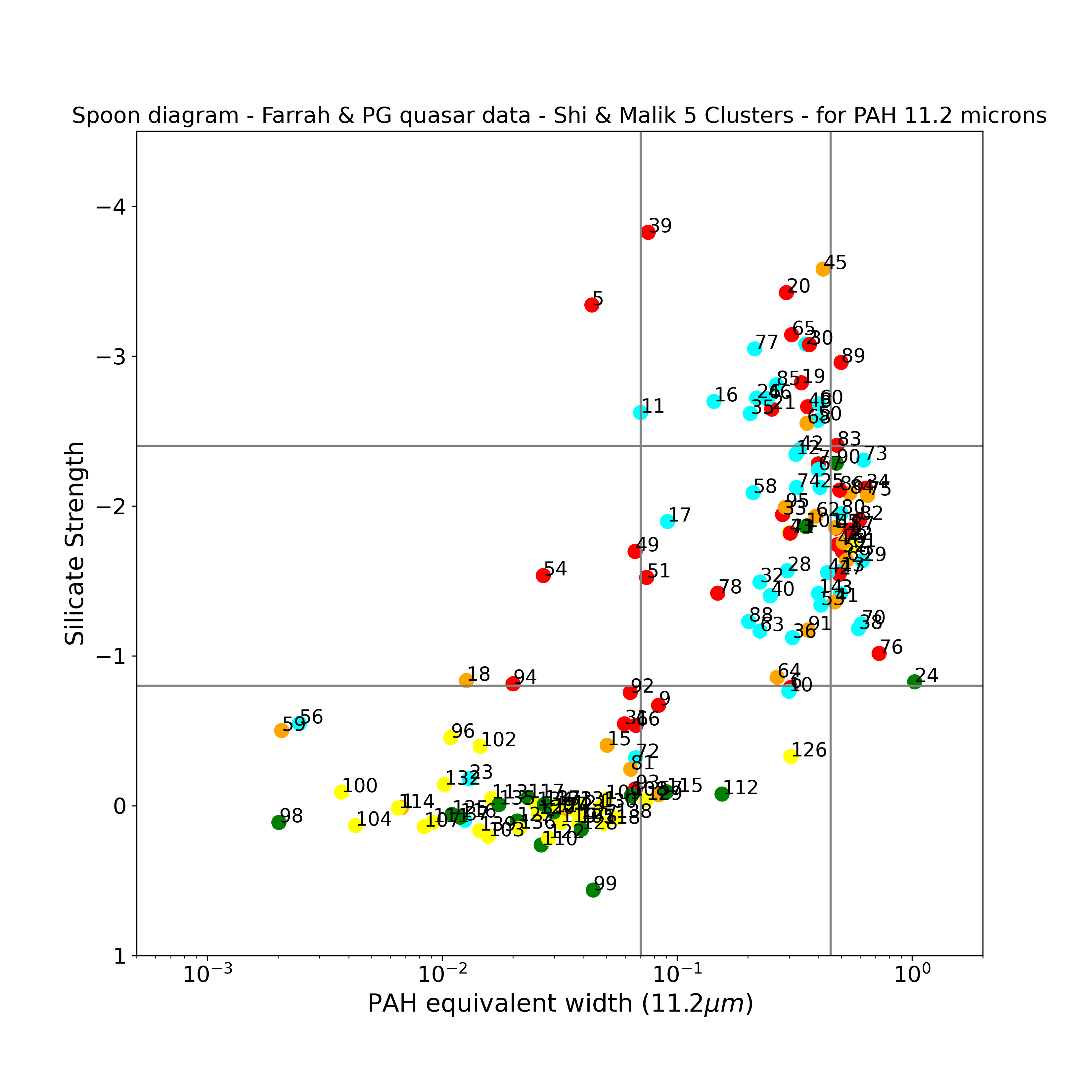}
\caption{ \; }\label{ClustersGraphGaussianPCA_PAH_11.2}
\end{subfigure}
\end{multicols} 

\caption{Fork diagrams, as introduced in \protect\citet{Spoon2007}, with galaxies color coded based on the clusters detected using the spectral clustering algorithm of \protect\citet{Shi_Malik2000}. The galaxies are distributed based on their silicate dust absorption feature (y-axis) and their (a) $6.2 \mu m$ and (b) $11.2 \mu m$ PAH emission features (x-axis). The diagram is separated into the nine regions (1A-3C) proposed by \protect\citet{Spoon2007}. The yellow and green-colored communities showcase features correlated to AGN-dominated quasars, whereas the red and cyan-colored communities showcase features of starburst-dominated galaxies. The orange-colored community indicates a mix of both starburst \& AGN-dominated galaxies.}\label{Spoon_Shi_Malik}

\end{figure*}

Following the \textit{fork} diagnostic diagram introduced by \citet{Spoon2007}, we can examine the clustering algorithm detected communities plotted on a 2-dimensional space, where the $x$-axis represents the PAH emission feature and the $y$-axis represents the silicate dust absorption feature. In Figures \ref{ClustersGraphGaussianPCA_PAH_6.2} and \ref{ClustersGraphGaussianPCA_PAH_11.2} we present the \textit{fork} diagrams with the galaxies color-coded based on the clusters identified by the implementation of the \citet{Shi_Malik2000} spectral graph clustering algorithm (Figure \ref{Shi_Malik_communities}). In these diagrams the galaxies are distributed on the y-axis based on the value of their silicate dust absorption feature and on the x-axis based on the value of their PAH equivalent width at $6.2 \mu m$ (Figure \ref{ClustersGraphGaussianPCA_PAH_6.2}) and $11.2 \mu m$ (Figure \ref{ClustersGraphGaussianPCA_PAH_11.2}). As expected, the starburst galaxies of the red and cyan-colored clusters (which correspond to LINERs and $H_{II}$ type galaxies) mostly correlate to higher PAH emission and silicate absorption features, whereas the AGN-dominated luminous quasars of the yellow and green-colored clusters correlate to very low PAH emission and extinct silicate absorption features. The orange-colored cluster galaxies are spread throughout the diagram, corresponding to a mix of both starburst and AGN-dominated galaxies. \par

By comparing the distribution of galaxies within the nine regions (1A-3C) presented in \citet{Spoon2007}, where numbers $1-3$ correspond to three separate regions of increasing silicate absorption (in ascending order) and letters \textit{A-C} correspond to three regions of increasing PAH emission (alphabetically), we observe that our classification scheme (through clustering analysis) suggests a distribution of the {\em five} classes of galaxies into the nine regions proposed by \citet{Spoon2007} in the following way:
\begin{itemize}
    \item The \textit{red} cluster: mainly in regions 3A-3B and 2A-2C.
    \item The \textit{cyan} cluster: mainly in regions 3A-3B and 2B-2C.
    \item The \textit{orange} cluster: in regions 3B, 2A-2C and 1A-1B.
    \item The \textit{green} cluster: mainly in regions 2B-2C and 1A-1B.
    \item The \textit{yellow} cluster: contained mainly within regions 1A-1B, with a few outliers also present in region 2B.
\end{itemize}

Overall, the performance of our clustering analysis, and the implementation of the spectral graph clustering algorithm of \citet{Shi_Malik2000} in particular, is in agreement with the distinct physical properties of the different regions, as proposed by \citet{Spoon2007}. Although there is an observed overlap of galaxy clusters within some of the regions, the containment of most of the galaxies of each cluster within specific regions is an important indicator of the good performance of the algorithm.

\vspace{3mm}

\subsubsection{Archetypical galaxies of each cluster}

By considering the observed features and physical properties of specific \textit{archetypical} galaxies for each of the detected clusters, we can further examine the evolution of power sources between clusters. The selection of these archetypical galaxies is initially based on the calculated \textit{local clustering coefficient} of individual galaxies (e.g. the probability that two neighbors of a vertex are connected), compared to the \textit{global clustering coefficient} of our graph (which is based on the ratio of connected triplets over all triangles of nodes in the graph). The chosen archetypical galaxies of each cluster display a higher clustering coefficient compared to the rest of the graph, meaning that they are more highly connected compared to the rest of the graph, and thus they can be interpreted as indicative for each local cluster. Additionally, where available, we select well-known galaxies as archetypical galaxies, based on established literature, namely \textit{Arp 220} (7) and \textit{3C 273} (96) for the red and yellow clusters respectively.

Beginning from the red cluster, \textit{Arp 220} (7) is a well known starburst galaxy, classified as a LINER type galaxy, displaying both high silicate absorption and high PAH emission features, as shown in Figures \ref{ClustersGraphGaussianPCA_PAH_6.2} and \ref{ClustersGraphGaussianPCA_PAH_11.2}. This indicates a high dust content around its power source, as well as high star-forming activity. Arp 220 also shows low nuclear separation, between 0.1-6 kpc (Figure \ref{Nuclear_separation_graph}). 

Considering the cyan cluster, \textit{IRAS 17068+4027} (74) is an $H_{II}$ type starburst galaxy (Figure \ref{Optical_Spectral_Type_graph}), which displays high nuclear separation (Figure \ref{Nuclear_separation_graph}) and moderately high silicate absorption and PAH emission features (Figures \ref{ClustersGraphGaussianPCA_PAH_6.2} and \ref{ClustersGraphGaussianPCA_PAH_11.2}). These properties are consistent with the features observed in the cyan cluster's mean signal, in Figure \ref{Mean_Signals_graph}. 

Regarding the orange cluster, it is a mixed cluster which presents moderate features of silicate absorption and PAH emission, thus we consider two archetypical galaxies: \textit{IRAS 00275-2859} (15) and \textit{IRAS 17179+5444} (75). The former is a Seyfert-I type galaxy which displays low silicate absorption and PAH emission features (Figures \ref{ClustersGraphGaussianPCA_PAH_6.2} and \ref{ClustersGraphGaussianPCA_PAH_11.2}). The latter is a Seyfert-II type galaxy, which also displays moderate silicate absorption but high PAH emission features. This combination goes some way to explaining the moderate features observed in the mean signal of the cluster, as this cluster is mostly populated by Seyfert-II type galaxies, but also contains a mixture of starburst galaxies and quasars. Both galaxies also display a very low nuclear separation (single nucleus), as shown in Figure \ref{Nuclear_separation_graph}.

The quasar \textit{PG 0934+013} (116) could be considered as the archetypical galaxy for the green cluster, which (as expected) displays very low nuclear separation (single nucleus). This quasar displays almost no silicate absorption, as well as nearly extinct PAH emission features (Figures \ref{ClustersGraphGaussianPCA_PAH_6.2} and \ref{ClustersGraphGaussianPCA_PAH_11.2}).

\textit{3C 273} (96) is a well known quasar that could be considered as the archetypical galaxy of the yellow cluster, displaying all of the expected features for a quasar. Specifically, 3C 273 shows very low nuclear separation (Figure \ref{Nuclear_separation_graph}), as well as very low silicate absorption and highly extinct PAH emission features. This is in-line with the observed mean signal of the yellow cluster. As mentioned above, considering the fork diagrams of Figures \ref{ClustersGraphGaussianPCA_PAH_6.2} and \ref{ClustersGraphGaussianPCA_PAH_11.2}, most of the yellow cluster quasars show no silicate absorption (or even silicate emission) as well as very low (extinct) PAH emission features. These features may be the main cause for the distinction between the green and yellow clusters by our clustering algorithms.

\vspace{3mm}

\subsubsection{Graph Theoretic Evolutionary Path}

The partitioning of the similarity graph into clusters combined with the characterization of the SED communities in relation to the physical properties of the galaxies, also reveals some relation between the communities themselves. Through the partitioning obtained by all graph clustering algorithms tested, we can observe that - besides the agreement on the number of clusters detected - in all algorithms' outcomes, each one of the identified clusters has two other neighbouring clusters in the graph, forming all together a {\em path} connecting the neighbouring clusters. Moreover, PAH emission and silicate absorption features gradually increase/decrease between neighbouring clusters. Combining these results seems to suggest a possible quasi-linear evolutionary path for the power sources of ULIRGs. \par

Thus, the results of our graph theoretical methodology indicate a transitional evolution between different types of galaxies, with the pre-merger evolutionary phase being populated by mostly dust-obscured starburst-dominated galaxies (mainly LINERs and $H_{II}$ galaxies), while the coalescence phase being populated by a combination of starburst and obscured AGN-dominated galaxies (mostly Type-II and some Type-I Seyferts), and finally the post-merger phase being populated by mostly unobscured AGN-dominated galaxies and luminous quasars. While previous works (e.g. \citet{Farrah2009}) also suggest the possibility of the merger scenario following two distinct evolutionary paths, our results showcase a single evolutionary scenario for mergers. However, this result could be subject to further investigation. \par

\section{Conclusions}\label{Conclusions}
 
We have presented how graph theoretical and clustering analysis tools can be utilized for the classification of galaxies based on their mid-IR spectra, in order to extract meaningful information on the underlying mechanisms. We have showcased the successful implementation of KPCA on mid-IR spectra of ULIRGs for the construction of a similarity graph, as well as utilizing clustering algorithms in order to extract different communities, corresponding to separate evolutionary stages. Furthermore, the examination of the physical properties of the identified communities led to a physical interpretation supporting the evolutionary merger scenario for ULIRGs. \par

More specifically, we have presented how our graph theoretical methodology results in an evolutionary paradigm for ULIRGs supporting the merger scenario, by showcasing a quasi-linear transition of the dominant power sources of ULIRGs, from the dust-obscured starburst-dominated galaxies of the pre-merger phase towards the unobscured AGN-dominated luminous quasars of the post-merger phase. \par

Additionally, we have tested and compared the performance of multiple different clustering algorithms on astrophysical data, based on graph theoretical metrics as well as in comparison to the existing theoretical framework for galaxy evolution. \par 

Future works utilizing higher-resolution mid-infrared ULIRG spectra (obtained by future missions such as the James Webb Space Telescope), as well as extended catalogues of ULIRGs, can result in defining a much more concrete evolutionary paradigm for the merger scenario of ULIRGs. Furthermore, future examination of additional graph theoretical approaches in studying galaxy evolution, as well as potential comparisons of the mid-infrared spectra of ULIRGs across different redshift ranges, could offer important insights on the nature and physical mechanisms of galaxy evolution throughout the history of the Universe. \par

\vspace{5mm} 
\section*{Acknowledgements}

The authors acknowledge support from the research project \textit{EXCELLENCE/1216/0207 - "Graph Theoretical Tools in Sciences" (GRATOS)} funded by the Research \& Innovation Foundation (RIF) in Cyprus.

\section*{Data Availability Statement}

The observational data used for the purposes of this article are available in the \href{https://cassis.sirtf.com/atlas/welcome.shtml}{Combined Atlas of Sources with Spitzer IRS Spectra (CASSIS)} website. Some of the results of the graph theoretical analysis of the data, generated as part of the aforementioned research project \textit{EXCELLENCE/1216/0207 (GRATOS)} used in this article are available on the website of the project at \href{https://arc.euc.ac.cy/gratos/resources/}{Aristarchus Research Center website}.

\vspace{3mm}

\bibliographystyle{elsarticle-harv}
\bibliography{bibliography}

\newpage

\appendix

\section{}\label{Appendix_section_ShiMalik}

The spectral clustering algorithm proposed by \citet{Shi_Malik2000} consists of the following steps:
\begin{enumerate}
    \item[(1)] First, we compute the \textit{affinity matrix} using the  Gaussian Kernel. Then, we apply the $k$-nearest neighbour method (see Section (\ref{Similarity_Section}), to extract the sparse, weighted SED similarity graph $G(V, E,  \mathbf{W})$. 
     
    \item[(2)] We compute the \textit{unnormalized Laplacian} of the graph $G(V, E, \mathbf{W}$), $\mathbf{L}$:    $\mathbf{L} =\mathbf{L} -\mathbf{W} $. 
     \item[(3)]  We compute the {\em $k$ eigenvectors $x_1,\cdots x_k$} of $L$:  $\mathbf{L} \mathbf{X }= \lambda \mathbf{D} \mathbf{X} $, where $k$\footnote{Note   that this parameter $k$ is unrelated to the parameter $k$ used for the $k$-nearest neighbour method applied before. We use the same letter, since this is the typically used letter in  both methods ($k$-nearest-neighbours and spectral graph clustering).} is given as input,  $D$ = the diagonal matrix, with $D_{ii}$ = the sum of the $i$-th row of $\mathbf{W}$, $\mathbf{X} = x_1, x_2, ..., x_k$ is a $N \times k $ matrix containing the $k$-largest generalized eigenvectors of $L$ and $ \lambda$ is scalar, the {\em eigenvalue} of $L$. 
     \item[(4)] We apply the {\em $k$-means} clustering algorithm  on the $N \times k$  matrix $\mathbf{X} = [x_1 x_2 \cdots  x_k]$ and assign each node $ i\in [N]$  to the cluster in which row $i$ of matrix  $\mathbf{X}$ is assigned. 
\end{enumerate}

\vspace{15mm}

\section{}\label{Appendix_Clustering_Graphs}

The graphs resulting from the application of several community detection clustering algorithms are displayed in this section. \\ The clustering algorithms implemented are: \par
\begin{enumerate}
    \item Louvain algorithm of \citet{Blondel_2008}.
    \item Edge betweenness algorithm of \citet{Girvan2002}.
    \item Leiden algorithm of \citet{Leiden}.
    \item Leading eigenvector algorithm of \citet{Newman2004b}.
    \item Walktrap algorithm of \citet{Pascal2006}.
    \item Spectral clustering algorithm of \citet{Ng2001}.
    \item Spectral clustering algorithm of \citet{Shi_Malik2000}.
\end{enumerate}

\renewcommand{\thefigure}{B.\arabic{figure}}

\setcounter{figure}{0}
\begin{figure} 
\centering
\includegraphics[width=1\columnwidth]{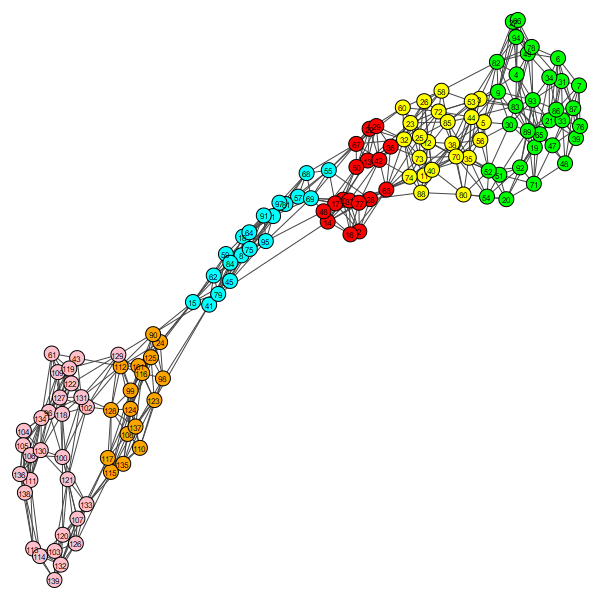}
\caption{Communities detected using the Louvain clustering algorithm of \protect\citet{Blondel_2008}. Six communities were detected with an optimal modularity of $0.6974$ and a max conductance of $0.2209$.}\label{Louvain_graph}
\end{figure}

\begin{figure} 
\centering
\includegraphics[width=1\columnwidth]{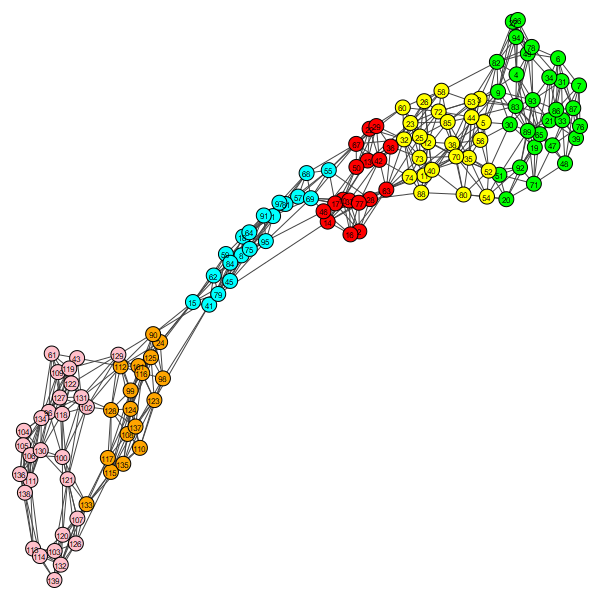}
\caption{Communities detected using the edge-betweenness clustering algorithm of \protect\citet{Girvan2002}. Six communities were detected with an optimal modularity of $0.6965$ and max conductance of $0.2209$.}\label{Girvan_Newman_graph}
\end{figure}

\begin{figure} 
\centering
\includegraphics[width=1\columnwidth]{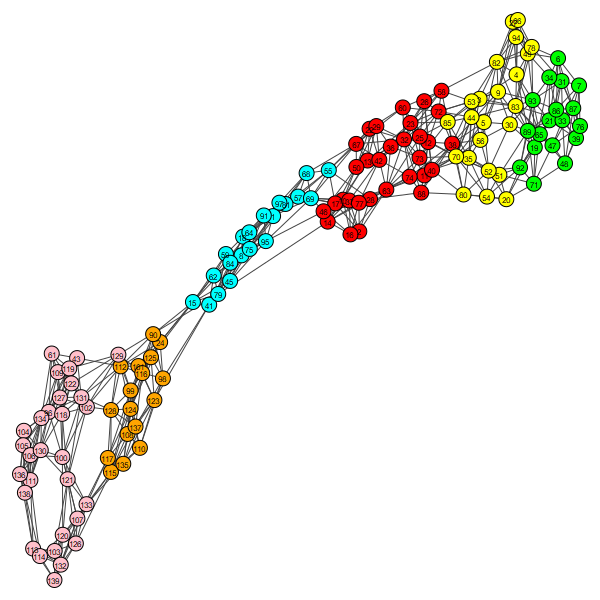}
\caption{Communities detected using the Leiden clustering algorithm of \protect\citet{Leiden}. Six communities were detected with a resolution parameter $= 1.0$, resulting in an optimal modularity of $0.6959$ and a conductance of $0.3069$.}\label{Leiden_graph}
\end{figure}

\begin{figure} 
\centering
\includegraphics[width=1\columnwidth]{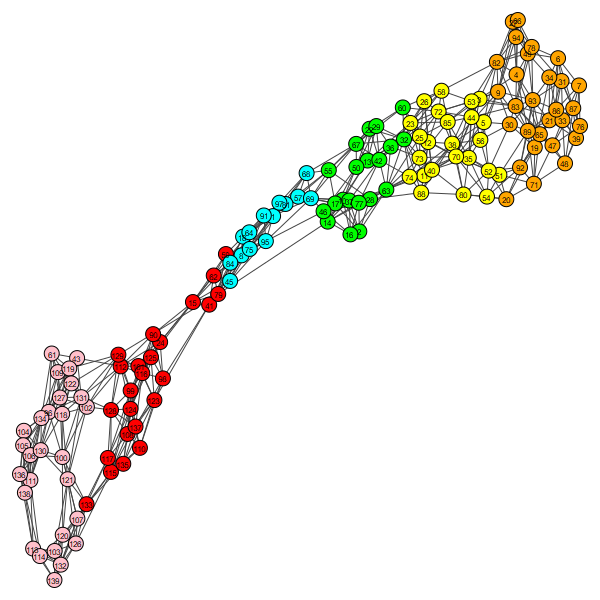}
\caption{Communities detected using the leading eigenvector clustering algorithm of \protect\citet{Newman2004b}. Six communities were detected with an optimal modularity of $0.6711$ and max conductance of $0.2288$.}\label{Newman_graph}
\end{figure}

\begin{figure} 
\centering
\includegraphics[width=1\columnwidth]{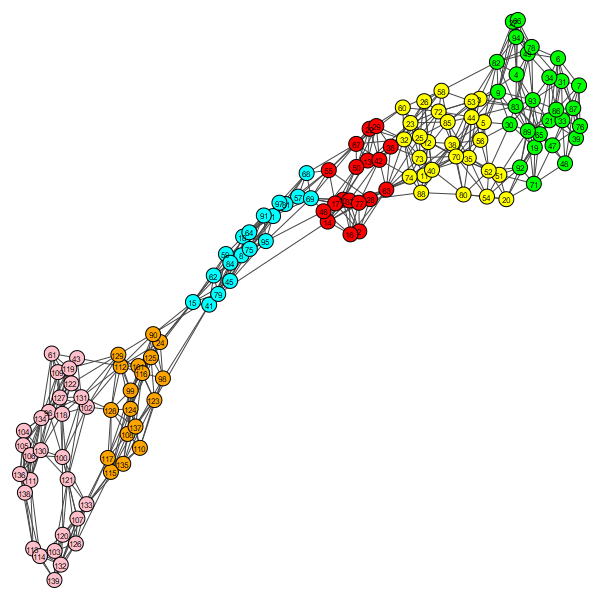}
\caption{Communities detected based on the random walks community detection algorithm of \protect\citet{Pascal2006}. Six communities were detected with an optimal modularity of $0.7012$ and max conductance of $0.2103$.}\label{Walktrap_graph}
\end{figure}

\begin{figure} 
\centering
\includegraphics[width=1\columnwidth]{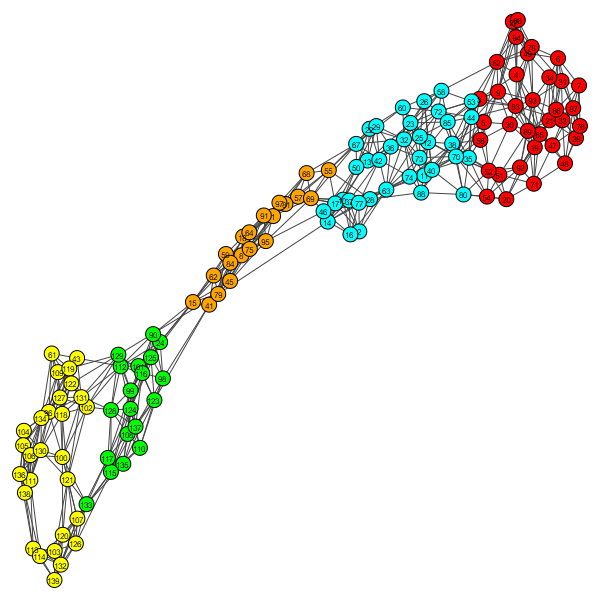}
\caption{Communities detected using the spectral clustering algorithm of \protect\citet{Ng2001}, with a preset of 5 clusters. The maximum conductance was calculated as $0.0526$ and the modularity score as $0.6841$.}\label{Andrew_Ng_graph}
\end{figure}

\begin{figure} 
\centering
\includegraphics[width=1\columnwidth]{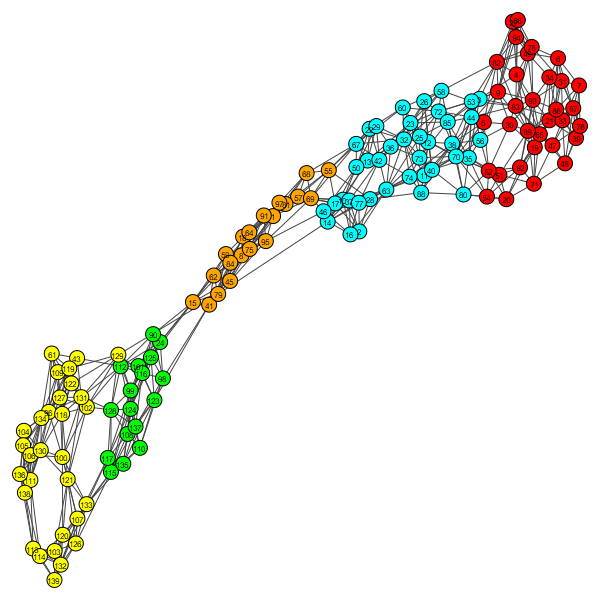}
\caption{Communities detected using the spectral clustering algorithm of \protect\citet{Shi_Malik2000}, with a preset of 5 clusters. The maximum conductance was calculated as $0.0588$ and the modularity score as $0.6899$.}\label{Shi_Malik_graph}
\end{figure}

\begin{figure} 
\centering
\includegraphics[width=1\columnwidth]{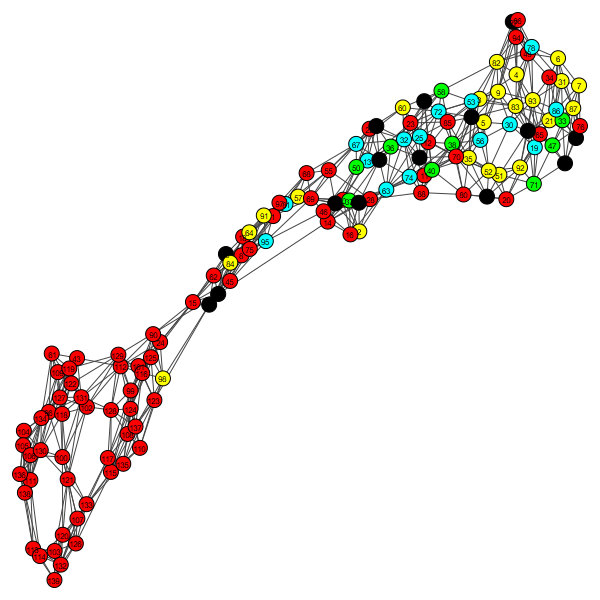}
\caption{Our graph colored based on the nuclear separation of ULIRGs, where data is available. Nuclear separation distance ranges are set (in kiloparsecs) similar to \protect\citet{Farrah2009} as: \\ Black: unavailable data, Cyan: $>12$ kpc, Green: $6-12$ kpc, Yellow: $0.1-6$ kpc, Red: $<0.1$ kpc (single nucleus).}\label{Nuclear_separation_graph}
\end{figure}

\begin{figure} 
\centering
\includegraphics[width=1\columnwidth]{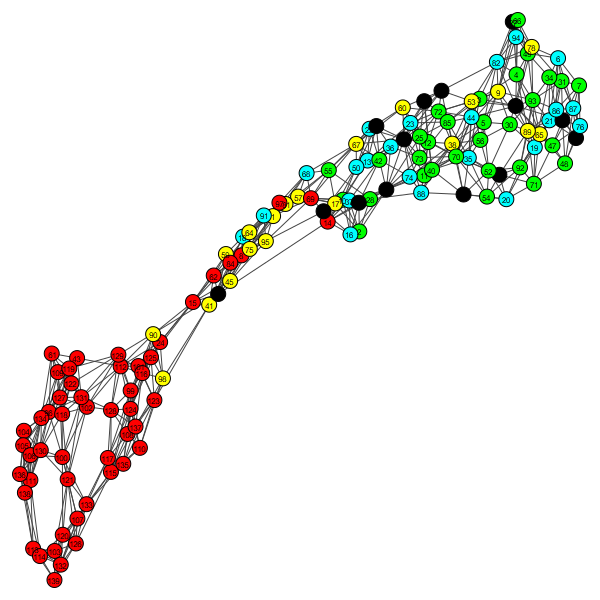}
\caption{Our graph colored based on the optical spectral type of ULIRGs, where data is available. Optical spectral type per color is taken similar to \protect\citet{Farrah2009} as: \\ Black = unavailable data, Cyan = $H_{II}$, Green = LINER, Yellow = Seyfert-2, Red = Seyfert-1/quasars.}\label{Optical_Spectral_Type_graph}
\end{figure}

\mbox{}
\clearpage
\newpage

\section{}\label{appendixIII}

\renewcommand{\thetable}{A.\arabic{table}}
\setcounter{table}{0}
\begin{table}[!htbp]
    \caption{Reference table of galaxy names and labels used in graphs.} \label{labels_table}
\begin{tabular}{|l| l| l| l| l| l| l| l|}
\hline
1  & IRAS~05189-2524   & 36 & IRAS~06009-7716 & 71  & IRAS~16300+1558      & 106 & PG~1151+117 \\ \hline
2  & IRAS~08572+3915   & 37 & IRAS~06035-7102 & 72  & IRAS~16334+4630      & 107 & PG~1307+085 \\ \hline
3  & IRAS~12112+0305   & 38 & IRAS~06206-6315 & 73  & IRAS~16576+3553      & 108 & PG~1309+355 \\ \hline
4  & IRAS~14348-1447   & 39 & IRAS~06301-7934 & 74  & IRAS~17068+4027      & 109 & PG~1402+261 \\ \hline
5  & IRAS~15250+3609   & 40 & IRAS~06361-6217 & 75  & IRAS~17179+5444      & 110 & PG~1501+106 \\ \hline
6  & IRAS~22491-1808   & 41 & IRAS~07145-2914 & 76  & IRAS~17208-0014      & 111 & PG~1535+547 \\ \hline
7  & Arp~220           & 42 & IRAS~07449+3350 & 77  & IRAS~17252+3659      & 112 & I Zw 1     \\ \hline
8  & Mrk~231           & 43 & IRAS~07598+6508 & 78  & IRAS~17463+5806      & 113 & PG~0049+171 \\ \hline
9  & Mrk~273           & 44 & IRAS~08208+3211 & 79  & IRAS~18030+0705      & 114 & PG~0921+525 \\ \hline
10 & UGC 5101         & 45 & IRAS~08559+1053 & 80  & IRAS~18443+7433      & 115 & PG~0923+129 \\ \hline
11 & IRAS~F00183-7111 & 46 & IRAS~09022-3615 & 81  & IRAS~19254-7245      & 116 & PG~0934+013 \\ \hline
12 & IRAS~00188-0856   & 47 & IRAS~09463+8141 & 82  & IRAS~19297-0406      & 117 & PG~1011-040 \\ \hline
13 & IRAS~00199-7426   & 48 & IRAS~10091+4704 & 83  & IRAS~19458+0944      & 118 & PG~1012+008 \\ \hline
14 & IRAS~00275-0044   & 49 & IRAS~10378+1109 & 84  & IRAS~20037-1547      & 119 & PG~1022+519 \\ \hline
15 & IRAS~00275-2859   & 50 & IRAS~10565+2448 & 85  & IRAS~20087-0308      & 120 & PG~1048+342 \\ \hline
16 & IRAS~00397-1312   & 51 & IRAS~11038+3217 & 86  & IRAS~20100-4156      & 121 & PG~1114+445 \\ \hline
17 & IRAS~00406-3127   & 52 & IRAS~11095-0238 & 87  & IRAS~20414-1651      & 122 & PG~1115+407 \\ \hline
18 & IRAS~01003-2238   & 53 & IRAS~11223-1244 & 88  & IRAS~20551-4250      & 123 & PG~1149-110 \\ \hline
19 & IRAS~01199-2307   & 54 & IRAS~11582+3020 & 89  & IRAS~21272+2514      & 124 & PG~1202+281 \\ \hline
20 & IRAS~01298-0744   & 55 & IRAS~12018+1941 & 90  & IRAS~23060+0505      & 125 & PG~1244+026 \\ \hline
21 & IRAS~01355-1814   & 56 & IRAS~12032+1707 & 91  & IRAS~23128-5919      & 126 & PG~1310-108 \\ \hline
22 & IRAS~01388-4618   & 57 & IRAS~12072-0444 & 92  & IRAS~23129+2548      & 127 & PG~1322+659 \\ \hline
23 & IRAS~01494-1845   & 58 & IRAS~12205+3356 & 93  & IRAS~23230-6926      & 128 & PG~1341+258 \\ \hline
24 & IRAS~02054+0835   & 59 & IRAS~12514+1027 & 94  & IRAS~23253-5415      & 129 & PG~1351+236 \\ \hline
25 & IRAS~02113-2937   & 60 & IRAS~13120-5453 & 95  & IRAS~23498+2423      & 130 & PG~1404+226 \\ \hline
26 & IRAS~02115+0226   & 61 & IRAS~13218+0552 & 96  & 3C~273               & 131 & PG~1415+451 \\ \hline
27 & IRAS~02455-2220   & 62 & IRAS~13342+3932 & 97  & Mrk 1014            & 132 & PG~1416-129 \\ \hline
28 & IRAS~02530+0211   & 63 & IRAS~13352+6402 & 98  & Mrk 463E            & 133 & PG~1448+273 \\ \hline
29 & IRAS~03000-2719   & 64 & IRAS~13451+1232 & 99  & PG~1119+120          & 134 & PG~1519+226 \\ \hline
30 & IRAS~03158+4227   & 65 & IRAS~14070+0525 & 100 & PG~1211+143          & 135 & PG~1534+580 \\ \hline
31 & IRAS~03521+0028   & 66 & IRAS~14378-3651 & 101 & PG~1351+640          & 136 & PG~1552+085 \\ \hline
32 & IRAS~03538-6432   & 67 & IRAS~15001+1433 & 102 & PG~2130+099          & 137 & PG~1612+261 \\ \hline
33 & IRAS~04114-5117   & 68 & IRAS~15206+3342 & 103 & PG~0052+251          & 138 & PG~2209+184 \\ \hline
34 & IRAS~04313-1649   & 69 & IRAS~15462-0450 & 104 & PG~0804+761          & 139 & PG~2304+042 \\ \hline
35 & IRAS~04384-4848   & 70 & IRAS~16090-0139 & 105 & PG~1116+215          &     &            \\ \hline
\end{tabular}
\end{table}

\mbox{}
\clearpage
\newpage

\section{}\label{appendixIV}

\renewcommand{\thetable}{A.\arabic{table}}
\setcounter{table}{1}
\begin{table}[!htbp]
\centering
\parbox{15cm}{\caption{Table of galaxy names and the calculated values for PAH emission equivalent widths at $6.2\mu m$ and $11.2\mu m$ and the silicate absorption strength at $9.7\mu m$ of our sample.}\label{PAH_Si_table}}
\resizebox{0.8\textwidth}{!}{%
\begin{tabular}{|c|c|c|c|c|c|c|c|}
\hline
\multicolumn{1}{|c|}{\textbf{Name}} &
  \multicolumn{1}{c|}{\textbf{PAH 6.2$\mu m$ ew}} &
  \multicolumn{1}{c|}{\textbf{PAH 11.2$\mu m$ ew}} &
  \multicolumn{1}{c|}{\textbf{Si 9.7$\mu m$}} &
  \multicolumn{1}{c|}{\textbf{Name}} &
  \multicolumn{1}{c|}{\textbf{PAH 6.2$\mu m$ ew}} &
  \multicolumn{1}{c|}{\textbf{PAH 11.2$\mu m$ ew}} &
  \multicolumn{1}{c|}{\textbf{Si 9.7$\mu m$}} \\ \hline
3C 273           & 0.01 & 0.01 & 0.01  & IRAS 17068+4027      & 0.21 & 0.30 & -1.82 \\ \hline
Arp 220          & 0.37 & 0.35 & -3.08 & IRAS 17179+5444      & 0.16 & 0.07 & -0.32 \\ \hline
IRAS 01388-4618  & 0.70 & 0.50 & -1.41 & IRAS 17208-0014      & 0.63 & 0.62 & -2.31 \\ \hline
IRAS 02113-2937  & 0.76 & 0.48 & -1.74 & IRAS 17252+3659      & 0.41 & 0.32 & -2.12 \\ \hline
IRAS 02530+0211  & 0.04 & 0.04 & -3.34 & IRAS 17463+5806      & 0.37 & 0.65 & -2.07 \\ \hline
IRAS 03000-2719  & 0.52 & 0.30 & -0.79 & IRAS 18030+0705      & 0.91 & 0.72 & -1.02 \\ \hline
IRAS 04384-4848  & 0.52 & 0.40 & -2.28 & IRAS 18443+7433      & 0.14 & 0.21 & -3.05 \\ \hline
IRAS 06009-7716  & 0.85 & 0.55 & -1.77 & IRAS 19254-7245south & 0.09 & 0.15 & -1.42 \\ \hline
IRAS 07145-2914  & 0.06 & 0.08 & -0.67 & IRAS 19297-0406      & 0.64 & 0.51 & -1.75 \\ \hline
IRAS 08559+1053  & 0.32 & 0.30 & -0.76 & IRAS 19458+0944      & 0.75 & 0.50 & -1.95 \\ \hline
IRAS F00183-7111 & 0.04 & 0.07 & -2.62 & IRAS 20037-1547      & 0.08 & 0.06 & -0.24 \\ \hline
IRAS 00188-0856  & 0.09 & 0.32 & -2.34 & IRAS 20087-0308      & 0.55 & 0.60 & -1.91 \\ \hline
IRAS 00199-7426  & 0.65 & 0.50 & -1.56 & IRAS 20100-4156      & 0.20 & 0.48 & -2.40 \\ \hline
IRAS 00275-0044  & 0.37 & 0.40 & -1.42 & IRAS 20414-1651      & 0.86 & 0.54 & -2.08 \\ \hline
IRAS 00275-2859  & 0.03 & 0.05 & -0.40 & IRAS 20551-4250      & 0.15 & 0.26 & -2.81 \\ \hline
IRAS 00397-1312  & 0.03 & 0.14 & -2.70 & IRAS 21272+2514      & 0.30 & 0.49 & -2.10 \\ \hline
IRAS 00406-3127  & 0.07 & 0.09 & -1.90 & IRAS 22491-1808      & 0.66 & 0.55 & -1.84 \\ \hline
IRAS 01003-2238  & 0.06 & 0.01 & -0.83 & IRAS 23128-5919      & 0.45 & 0.20 & -1.23 \\ \hline
IRAS 01199-2307  & 0.21 & 0.34 & -2.82 & IRAS 23129+2548      & 0.10 & 0.50 & -2.96 \\ \hline
IRAS 01298-0744  & 0.01 & 0.29 & -3.42 & IRAS 23230-6926      & 0.50 & 0.48 & -2.28 \\ \hline
IRAS 01355-1814  & 0.32 & 0.25 & -2.64 & IRAS 23253-5415      & 0.31 & 0.36 & -1.17 \\ \hline
IRAS 01494-1845  & 0.82 & 0.54 & -1.77 & IRAS 23498+2423      & 0.09 & 0.06 & -0.75 \\ \hline
IRAS 02054+0835  & 0.01 & 0.01 & -0.18 & Mrk 1014             & 0.08 & 0.07 & -0.11 \\ \hline
IRAS 02115+0226  & 0.36 & 1.03 & -0.83 & Mrk 231              & 0.03 & 0.02 & -0.81 \\ \hline
IRAS 02455-2220  & 0.68 & 0.41 & -2.12 & Mrk 273              & 0.30 & 0.29 & -1.99 \\ \hline
IRAS 03158+4227  & 0.14 & 0.22 & -2.72 & Mrk 463E             & 0.01 & 0.01 & -0.45 \\ \hline
IRAS 03521+0028  & 0.70 & 0.49 & -1.54 & PG 1119+120          & 0.01 & 0.03 & 0.03  \\ \hline
IRAS 03538-6432  & 0.26 & 0.29 & -1.57 & PG 1211+143          & 0.01 & 0.00 & 0.11  \\ \hline
IRAS 04114-5117  & 0.71 & 0.61 & -1.63 & PG 1351+640          & 0.03 & 0.04 & 0.56  \\ \hline
IRAS 04313-1649  & 0.22 & 0.37 & -3.08 & PG 2130+099          & 0.01 & 0.00 & -0.09 \\ \hline
IRAS 05189-2524  & 0.04 & 0.06 & -0.54 & UGC 5101             & 0.35 & 0.35 & -1.86 \\ \hline
IRAS 06035-7102  & 0.14 & 0.23 & -1.49 & WIR-IRAS 23060+0505  & 0.02 & 0.01 & -0.40 \\ \hline
IRAS 06206-6315  & 0.25 & 0.28 & -1.94 & I Zw 1               & 0.00 & 0.02 & 0.20  \\ \hline
IRAS 06301-7934  & 0.39 & 0.63 & -2.12 & PG 0049+171          & 0.01 & 0.00 & 0.13  \\ \hline
IRAS 06361-6217  & 0.11 & 0.20 & -2.62 & PG0052+251           & 0.00 & 0.04 & 0.11  \\ \hline
IRAS 07449+3350  & 0.35 & 0.31 & -1.12 & PG0804+761           & 0.01 & 0.00 & 0.34  \\ \hline
IRAS 07598+6508  & 0.02 & 0.01 & 0.10  & PG 0921+525          & 0.00 & 0.01 & 0.14  \\ \hline
IRAS 08208+3211  & 0.70 & 0.59 & -1.18 & PG 0923+129          & 0.04 & 0.06 & -0.07 \\ \hline
IRAS 08572+3915  & 0.04 & 0.08 & -3.82 & PG 0934+013          & 0.04 & 0.05 & -0.04 \\ \hline
IRAS 09022-3615  & 0.20 & 0.25 & -1.40 & PG 1011-040          & 0.01 & 0.03 & 0.26  \\ \hline
IRAS 09463+8141  & 0.51 & 0.47 & -1.36 & PG 1012+008          & 0.01 & 0.01 & 0.11  \\ \hline
IRAS 10091+4704  & 0.07 & 0.33 & -2.37 & PG 1022+519          & 0.09 & 0.16 & -0.08 \\ \hline
IRAS 10378+1109  & 0.10 & 0.30 & -1.82 & PG 1048+342          & 0.03 & 0.02 & -0.05 \\ \hline
IRAS 10565+2448  & 0.70 & 0.44 & -1.55 & PG 1114+445          & 0.00 & 0.01 & 0.02  \\ \hline
IRAS 11038+3217  & 0.17 & 0.42 & -3.58 & PG 1115+407          & 0.06 & 0.09 & -0.09 \\ \hline
IRAS 11095-0238  & 0.04 & 0.24 & -2.72 & PG 1116+215          & 0.00 & 0.01 & 0.08  \\ \hline
IRAS 11223-1244  & 0.33 & 0.48 & -1.85 & PG 1149-110          & 0.01 & 0.02 & -0.05 \\ \hline
IRAS 11582+3020  & 0.11 & 0.36 & -2.66 & PG 1151+117          & 0.26 & 0.05 & 0.12  \\ \hline
IRAS 12018+1941  & 0.17 & 0.07 & -1.70 & PG 1202+281          & 0.02 & 0.03 & 0.11  \\ \hline
IRAS 12032+1707  & 0.01 & 0.40 & -2.57 & PG 1244+026          & 0.03 & 0.03 & 0.05  \\ \hline
IRAS 12072-0444  & 0.15 & 0.07 & -1.52 & PG 1307+085          & 0.11 & 0.04 & 0.02  \\ \hline
IRAS 12112+0305  & 0.61 & 0.51 & -1.70 & PG 1309+355          & 0.01 & 0.03 & 0.22  \\ \hline
IRAS 12205+3356  & 0.65 & 0.41 & -1.34 & PG 1310-108          & 0.00 & 0.02 & 0.10  \\ \hline
IRAS 12514+1027  & 0.00 & 0.03 & -1.53 & PG 1322+659          & 0.03 & 0.03 & 0.04  \\ \hline
IRAS 13120-5453  & 0.71 & 0.47 & -1.85 & PG 1341+258          & 0.02 & 0.01 & 0.06  \\ \hline
IRAS 13218+0552  & 0.00 & 0.00 & -0.55 & PG 1351+236          & 0.26 & 0.30 & -0.33 \\ \hline
IRAS 13342+3932  & 0.16 & 0.08 & -0.07 & PG 1402+261          & 0.01 & 0.04 & 0.09  \\ \hline
IRAS 13352+6402  & 0.12 & 0.21 & -2.09 & PG 1404+226          & 0.03 & 0.04 & 0.16  \\ \hline
IRAS 13451+1232  & 0.01 & 0.00 & -0.50 & PG 1415+451          & 0.01 & 0.07 & -0.04 \\ \hline
IRAS 14070+0525  & 0.12 & 0.40 & -2.68 & PG 1416-129          & 0.03 & 0.05 & 0.01  \\ \hline
IRAS 14348-1447  & 0.49 & 0.56 & -1.73 & PG 1448+273          & 0.02 & 0.04 & -0.00 \\ \hline
IRAS 14378-3651  & 0.74 & 0.39 & -1.93 & PG 1501+106          & 0.01 & 0.01 & -0.14 \\ \hline
IRAS 15001+1433  & 0.29 & 0.23 & -1.16 & PG 1519+226          & 0.00 & 0.03 & -0.00 \\ \hline
IRAS 15206+3342  & 0.46 & 0.27 & -0.86 & PG 1534+580          & 0.01 & 0.03 & 0.00  \\ \hline
IRAS 15250+3609  & 0.03 & 0.31 & -3.14 & PG 1535+547          & 0.00 & 0.02 & -0.01 \\ \hline
IRAS 15462-0450  & 0.09 & 0.07 & -0.53 & PG 1552+085          & 0.02 & 0.02 & 0.15  \\ \hline
IRAS 16090-0139  & 0.13 & 0.40 & -2.24 & PG 1612+261          & 0.00 & 0.03 & -0.01 \\ \hline
IRAS 16300+1558  & 0.14 & 0.36 & -2.55 & PG 2209+184          & 0.04 & 0.05 & 0.08  \\ \hline
IRAS 16334+4630  & 0.60 & 0.53 & -1.64 & PG 2304+042          & 0.01 & 0.01 & 0.17  \\ \hline
IRAS 16576+3553  & 0.48 & 0.61 & -1.21 &                      &      &      &       \\ \hline
\end{tabular}%
}
\end{table}

\end{document}